\documentclass[12pt,journal,onecolumn]{IEEEtran}
\usepackage{amsmath,amssymb,threeparttable,placeins}
\usepackage{lipsum}
\usepackage{bm}
\usepackage{multicol}
\usepackage{array}

\usepackage[mathscr]{euscript}
\usepackage[dvips]{graphicx}
\usepackage[usenames]{color}
\usepackage{amsfonts}
\usepackage{latexsym}
\usepackage{subfigure}
\usepackage[format=hang]{subfig}
\usepackage[justification=centering]{caption}
\usepackage{floatrow}
\usepackage{multirow}
\usepackage{graphicx}
\usepackage{epstopdf}
\usepackage[english]{babel}
\DeclareMathAlphabet{\mathpzc}{OT1}{pzc}{m}{it}

\newtheorem{theorem}{{{\textbf{\textit{Theorem}}}}}

\newtheorem{lemma}{{{\textbf{\textit{Lemma}}}}}
\newtheorem{corollary}{{{{\textbf{\textit{Corollary}}}}}}
\newtheorem{property}{{{\textbf{\textit{Property}}}}}
\newtheorem{definition}{\textbf{\textit{Definition}}}

\newtheorem{remark}{{{\textbf{\textit{Remark}}}}}

\newtheorem{example}{{{\textbf{\textit{Example}}}}}
\newtheorem{result}{{{\textbf{\textit{Result}}}}}

\begin{document}

\title{New Optimal $Z$-Complementary Code Sets from Matrices of Polynomials}
\author{Shibsankar Das,~%\IEEEmembership{Student Member,~IEEE},
        Udaya Parampalli,~%~\IEEEmembership{Senior Member,~IEEE},
        Sudhan Majhi,~%\IEEEmembership{Senior Member,~IEEE},\\
        and~Zilong Liu %~\IEEEmembership{Member,~IEEE}
\thanks{S. Das is a visiting Ph.D. student with  the School of Computing and Information Systems, The University of Melbourne, Melbourne, VIC 3053, Australia, on academic leave from the Department of Mathematics, IIT Patna, Bihar 801103, India (e-mail: shibsankar.pma15@iitp.ac.in). U. Parampalli is with the School of Computing and Information Systems, The University of Melbourne, Melbourne, VIC 3053, Australia (e-mail: udaya@unimelb.edu.au). S. Majhi is with the Department of Electrical Engineering, IIT Patna, Bihar 801103, India (e-mail: smajhi@iitp.ac.in). Z. Liu is with the Institute for Communication Systems, 5G Innovation Centre, University of Surrey, Guildford, GU2 7XH, U.K. (e-mail: zilong.liu@surrey.ac.uk).

A part of this work has been submitted to IEEE International Workshop on Signal Design and its Applications 2019 (IWSDA'19).
}
}
 \maketitle
 
\begin{abstract}
The concept of paraunitary (PU) matrices arose in the early 1990s in the study of multi-rate filter banks. So far, these matrices have found wide applications in cryptography, digital signal processing, and wireless communications. Existing PU matrices are subject to certain constraints on their existence and hence their availability is not guaranteed in practice. Motivated by this, for the first time, we introduce a novel concept, called $Z$-paraunitary (ZPU) matrix, whose orthogonality is defined over a matrix of  polynomials with identical degree not necessarily taking the maximum value. We show that there exists an equivalence between a ZPU matrix and a $Z$-complementary code set when the latter is expressed as a matrix with polynomial entries. Furthermore, we investigate some important properties of ZPU matrices, which are useful for the extension of matrix sizes and sequence lengths.  Finally, we propose a unifying construction framework for optimal ZPU matrices which includes existing PU matrices as a special case.
\end{abstract}
\begin{IEEEkeywords}
Paraunitary Matrices, $Z$-Paraunitary Matrices, $Z$-Complementary Sequences, Zero Correlation Zone, Unimodular Sequences.
\end{IEEEkeywords}

\section{\textbf{Introduction}}
\label{sec:intro}
\subsection{\bf Background}
\label{subsec:background}
\IEEEPARstart{T}{HE} past few decades have witnessed significant advances on the study of matrices of polynomials.  A matrix of polynomials refers to a matrix whose entries are polynomials. One attractive feature of this class of matrices is that each can be expressed either as a matrix with polynomial entries or as a polynomial with matrix coefficients. For example, an $M\times K$ matrix $\textbf{X}(z)$ of polynomials over $z^{-1}$ can be expressed as follows:
\begin{equation}
\label{polynimial:matrix}
\textbf{X}(z)=\Big[x_{mk}(z)\Big]_{M\times K}=\sum_{l=0}^{L-1}\textbf{X}_l\cdot z^{-l},
\end{equation} 
where $x_{mk}(z)$ is the $(m,k)$-th element of $\mathbf{X}(z)$ which is a polynomial with degree $(L-1)$ over the indeterminate variable $z^{-1}$ and $\textbf{X}_l$ is an $M\times K$ matrix comprising coefficients of $z^{-l}$. In a $z$-transform, $z^{-1}$ represents a unit delay.

A paraunitary (PU) matrix  refers to a matrix of polynomials in the indeterminate variable $z^{-1}$ which is unitary on the unit circle. A constant PU matrix independent of $z^{-1}$ is a conventional  unitary matrix. In \cite{1993Vaidyanathan}, Vaidyanathan introduced the concept of PU matrices and showed that they play a central role for perfect reconstruction system in the theory of multi-rate filter-banks. Nowadays, PU matrices have found wide applications in numerous areas such as filter-bank theory \cite{1997Phoong}, \cite{2014Kofidis}, wavelets and multiwavelets \cite{1997Strang,1995Strang,1998Jiang}, control theory \cite{1980Kailath}, digital signal processing \cite{2007McWhirter}, cryptography \cite{2006Delgosha}, etc. In wireless communications, Phoong and Chang have shown that a binary PU precoded orthogonal frequency-division multiplexing (OFDM) system enjoys enhanced error probability performance than the uncoded OFDM systems \cite{2005Phoong}, \cite{2012PhoongMIMO_OFDM}. PU matrices have also been employed for precoding in code-division multiple access (CDMA) systems in \cite{1995Wornell}.

In recent years, there has been tremendous research interest on the design of complementary sequences from PU matrices and vice versa \cite{2012Budisin}\nocite{2016Wang}\nocite{2017APCCShibsankar}\nocite{2017Shibsankar}\nocite{2018Budishin}\nocite{2018SETAShibsankar}\nocite{2018MaSPL}-\cite{2018Shibsankar}. With the aid of $z$-transform, PU matrices turn out to be a powerful tool in simplifying the derivations of sequences with good correlation properties. In \cite{2012Budisin}, a compact formulation has been proposed for complementary sequence pairs\footnote{A complementary sequence pair is also known as a Golay complementary pair (GCP), a concept proposed by Golay in the late 1940s in his study of spectrometry \cite{1949golay}, \cite{1961golay}. A GCP,  consisting of two constituent sequences, exhibits zero aperiodic auto-correlation sums for all non-zero time-shifts. Every constituent sequence in a GCP is also called a Golay sequence.} (and sets) by using PU matrices. The applications of PU matrices have also been extended to the constructions of $q$-ary complementary sequence sets  \cite{2016Wang}, \cite{2018Budishinn}, and QAM complementary sequence sets \cite{2018Budishin}. It is worthy to mention that \cite{2016Wang} introduced the use of Butson-type Hadamard ($BH$) matrices for new PU matrices. By associating the coefficients of a PU matrix with multiple sequence matrices, it has been shown that there exists an equivalence between a PU matrix and a set of complete complementary codes (CCC) \cite{2017Shibsankar}, \cite{2018Shibsankar}. Constructions of CCCs through traditional sequence operations can be found in \cite{1988NSuehiro}\nocite{2007Marziani}\nocite{2008Rathinakumar}-\cite{2011Han}. \cite{2019ShibsankarTSP} presents a design of polyphase CCC with various sequence lengths based on direct sum of PU matrices. Very recently, new near-optimal zero correlation zone (ZCZ) sequence sets have been developed based on PU matrices \cite{2019ISITShibsankar}.
\begin{figure}
	\centering
	\includegraphics[width=0.4\textwidth]{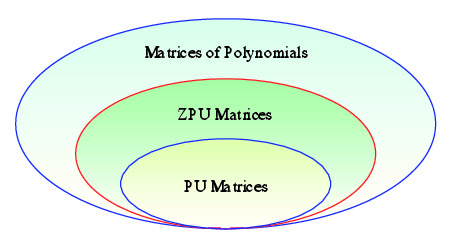}
	\caption{Relationship between ZPU matrix and PU matrix}
	\label{Figure:Z-PU:PU:relationship}
\end{figure}

Despite a wide range of applications of CCC in areas such as wireless communications \cite{2001Chen}, \cite{2015LiuFractionalDelayResilient} and information hiding \cite{2013Kojima}, \cite{2014Kojima}, CCC suffers from the small set size problem, i.e., the number of codes is upper bounded by the number of multi-channels, i.e., the number of constituent sequences in each code. To overcome this weakness, $Z$-complementary code sets (ZCCSs) have been proposed \cite{2007Fan_ZCSS}, \cite{2007Chen}, where $Z$ denotes the ZCZ width shared by all the codes. By definition, a ZCCS refers to a family of codes having zero auto- and cross-correlation properties within the ZCZ width $Z$. Significant research attention has been paid to ZCCS with two orthogonal channels. For a binary $Z$-complementary pair (ZCP), Fan \textit{et al.} conjectured in \cite{2007Fan_ZCSS} that the ZCZ width $Z$ satisfies $Z\leq L-2$, where $L$ ($\neq 2^{\alpha}10^{\beta}26^{\gamma}$) denotes the sequence length (even). \cite{2016Adhikary} proved that a binary periodic ZCP should also have even length. Li \textit{et al.} investigated the existence of binary ZCPs in \cite{2011Li}. In \cite{2014Liu_Z_CCS}, Liu \textit{et al.} proposed a construction of ZCPs with ZCZ width of $2^{m+1}$ and sequence length of $3\cdot 2^{m}$ by proper truncation of GCPs. Subsequently, Liu \textit{et al.} proposed a construction of optimal odd-length binary ZCPs, each displaying maximum ZCZ width and minimum out-of-zone aperiodic auto-correlation sums by applying insertion or deletion to certain carefully selected GCPs \cite{2014Liu_TIT}. Li \textit{et al.} proved that any ZCP can be written as a linear combination of a ZCP and its mutually orthogonal mate \cite{2016Li}. In \cite{2017Chen}, Chen proposed a direct construction of ZCPs with ZCZ width of $2^{m-2}+2^{\nu}$ and sequence length of $2^{m-1}+2^{\nu}$ based on generalized  Boolean functions (GBF), where $\nu=0,1,\cdots, m-2$. In \cite{2018AvikSPL}, Adhikary \textit{et al.} provided a construction of even-length binary ZCPs by insertion of concatenated odd-length binary ZCPs. Xie and Sun presented a construction of even-length binary ZCPs with ZCZ width of $2^{m+3}$ and sequence length of $7\cdot 2^{m+1}$ in \cite{2018Xie}. In \cite{2015Li_COML}, Li and Xu have shown that a ZCCS can be constructed from a Golay sequence with zero periodic auto-correlation zone (ZPACZ) \cite{2013Gong}. Direct constructions of ZCCSs based on GBF have been proposed in \cite{2018WuChenSPL} and \cite{2019PalashZCCS}, respectively. The zero correlation properties (within the ZCZ width) of ZCCS can enable interference-free multi-carrier CDMA (MC-CDMA) communications in quasi-synchronous channels \cite{2015LiuFractionalDelayResilient}, \cite{2008LiMCCDMAZCCS}, \cite{2018PalashIEEEAccess}. In addition, ZCCSs may be used for the peak-to-mean envelope power ratio (PMEPR) reduction in OFDM systems \cite{2018WuChenSPL}.

\subsection{\bf Motivations and Contributions}
 In \cite{1993Vaidyanathan}, it has been shown that any arbitrary PU matrix can be expressed as a product of unitary and diagonal matrices. This factorization is said to be an expanded product form of a PU matrix. For given number of phases, the existence of a PU matrix relies on the existence of unitary matrices of certain sizes. For example, a binary PU matrix of order $6\times 6$ does not exist since a $6 \times 6$ binary unitary matrix does not exist. In \cite{2005Phoong}, Phoong and Chang discussed the existence of binary PU matrices and postulated that: ``\textit{There are a number of open problems. For example, it is still unclear if there exist APU matrices\footnote{Here, a binary PU matrix is referred to as an antipodal PU (APU) matrix.} with odd length $\geq 3$.  All the above construction methods generate APU matrices of even lengths only. In addition, we do not know if there are APU matrices with dimensions of $4k+2$, for $k\geq 1$.}" Moreover, orthogonal analysis shows that an $M\times K$ PU matrix does not exist when $K>M$. These motivate us to investigate solutions to address the existence issues pertinent to PU matrices.

 \begin{figure}
	\centering
	\includegraphics[width=0.5\textwidth]{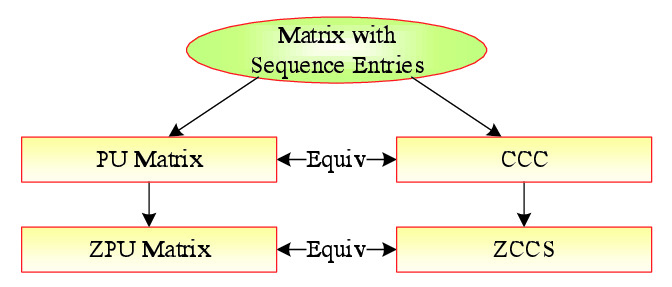}
	\caption{Relationship between ZPU matrix and ZCCS}
	\label{Figure:Z-PU:Z-CCS}
\end{figure}
From a sequence point of view, modern communication systems require very flexible choices of sequence lengths and set sizes without any sacrifice of the desired correlation properties. Existing ZCCS parameters are, however, mostly limited to powers of two. Driven by the success of polynomial matrices in the constructions of GCPs and CCC, it is interesting to exploit its application for the finding of new ZCCS. For example, a generic construction framework under matrices of polynomials for more flexible choices of ZCCS parameters remains largely open.

This paper presents a novel construction of ZCCSs described in a $z$-domain framework by introducing  the concept of $Z$-paraunitary (ZPU) matrices. The proposed ZPU concept includes the existing PU matrices as a special case. Fig. \ref{Figure:Z-PU:PU:relationship} portraits the relationship between ZPU matrix and PU matrix. The basic idea is to allow the range of time-shifts with zero correlations to be less than or equal to the sequence length, i.e., $Z\leq L$. We show that there exists a one-to-one correspondence between a ZPU matrix and a ZCCS when the sequences of the latter are expressed as polynomial entries of the former. We provide a diagram in Fig. \ref{Figure:Z-PU:Z-CCS} on the relationship between ZPU matrix and ZCCS as well as their individual evolutions. We study some important properties of ZPU matrices which are useful for the expansion of matrix sizes and sequence lengths. Based on these properties, we develop a unifying construction framework for optimal ZPU matrices, which includes existing PU matrices as a special case. Our main idea is to construct a ``fat" polynomial matrix (instead of a square one) by carefully expanding certain PU matrix in a way that concatenation or interleaving of CCC comes to interplay. We prove that such a polynomial matrix multiplied by its Hermitian will give rise to an identity matrix times the matrix energy, when all the polynomial terms with degrees not less than the ZCZ width are discarded in the calculation. We show that our proposed optimal ZPU matrices lead to optimal ZCCSs which meet their set size upper bound. The proposed construction framework not only simplifies the derivations of ZCCS constructions, but also offers more flexible choices of ZCCS parameters compared to the previously known ones.

\subsection{\bf Organization}
The remainder of the paper is organized as follows. In Section \ref{Relationship:previous:works:sec}, we review ZCCS parameters with the aid of a table summarizing the existing constructions.  In Section \ref{Preli:sec:back}, we present some basic definitions,  notations and a brief review of Butson-type Hadamard matrices. In Section \ref{Z-PU:Sec:Introduction}, we introduce the concept of ZPU matrices with examples and show the relationship between ZPU matrix and ZCCS. In Section \ref{Unifying:Framework:sec:pro}, we propose a unifying  construction framework for optimal ZPU matrices.  Finally, we conclude our work in Section \ref{Z-PU:sec:conclusion}.

 \section{\textbf{Brief Review on Existing ZCCS Parameters}}
\label{Relationship:previous:works:sec}
In this section, we will first briefly review previously known ZCCS parameters. Then, we will compare the parameters of our proposed ZCCSs with that of the previous works through a table.

So far, there are four types of construction methods for ZCCSs: the first type is based on GBFs \cite{2017Chen}, \cite{2018WuChenSPL}, \cite{2019PalashZCCS}, the second based on seed ZCPs  \cite{2007Fan_ZCSS}, \cite{2016Li}, the third based on GCPs \cite{2014Liu_Z_CCS}, \cite{2014Liu_TIT}, \cite{2018AvikSPL}, \cite{2018Xie}, and the fourth based on ZPACZ Golay sequences \cite{2015Li_COML}. Most these  algorithms have been concerned with ZCPs.  In fact, \cite{2015Li_COML}, \cite{2018WuChenSPL}, and \cite{2019PalashZCCS} studied ZCCSs with constituent sequences of more than two.

Specifically, in \cite{2015Li_COML}, Li and Xu proposed a construction for ZCCSs based on Golay sequences with ZPACZ. Their set size, flock size, ZCZ width and sequence length are $rZ$, $L$, $s$, and $rs$, respectively, where $L$ is the length of a Golay sequence with ZPACZ $Z$ and $s|Z$ for some positive integers $r$ and $s$. The method in \cite{2015Li_COML} can generate optimal ZCCSs only when Golay sequences with zero periodic auto-correlation functions (i.e., perfect sequences\footnote{A sequence is referred to as a perfect sequence if the periodic auto-correlation sidelobes are all zero \cite{1995Mow}.}) are used. The parameters of the Wu-Chen ZCCSs \cite{2018WuChenSPL} are limited to powers of two. The same can be said for the ZCCS construction proposed in \cite{2019PalashZCCS}. In Table \ref{Table:Relationship with Previous Works}, we compare the existing ZCCS parameters with our proposed ones. Table \ref{Table:Relationship with Previous Works} shows that our proposed construction framework offers more ZCCSs which may not be generated by previous construction methods. For instance, an optimal $(6,9)$-ZCCS$^{18}_{3}$ (see Table \ref{Table:Z-PU:M3:K6:Z9:L18}) may not be generated by the previous construction methods.

 \begin{center}
\captionof{table}{Summary of Existing ZCCS Parameters}  \label{Table:Relationship with Previous Works}
	\resizebox{\linewidth}{!}{
		\renewcommand{\arraystretch}{1.51}
\begin{tabular}{|l|l|l|l|l|l|l|l|l|l|l|}
\hline
\multicolumn{1}{|c|}{Reference} & \multicolumn{1}{c|}{Based On} & \multicolumn{1}{c|}{Phase} & \multicolumn{1}{c|}{Set Size} & \multicolumn{1}{c|}{Flock Size} & \multicolumn{1}{c|}{ZCZ Width} & \multicolumn{1}{c|}{Length} & \multicolumn{1}{c|}{Constraints} & \multicolumn{1}{c|}{Optimality} \\
\hline
\hline
\multicolumn{1}{|c|}{Li \cite{2015Li_COML}} & \multicolumn{1}{c|}{Length-$L$ ZPACZ Sequence} & \multicolumn{1}{c|}{$q$} & \multicolumn{1}{c|}{$rZ$} & \multicolumn{1}{c|}{$L$} & \multicolumn{1}{c|}{$s$} & \multicolumn{1}{c|}{$rs$} & \multicolumn{1}{c|}{$r,s,q\geq 2, s|Z$; $2|q$} & \multicolumn{1}{c|}{Not Optimal} \\
\hline
\multicolumn{1}{|c|}{Wu \cite{2018WuChenSPL}} & \multicolumn{1}{c|}{Boolean Functions} & \multicolumn{1}{c|}{$q$} & \multicolumn{1}{c|}{$2^{k+\nu}$} & \multicolumn{1}{c|}{$2^k$} & \multicolumn{1}{c|}{$2^{m-\nu}$} & \multicolumn{1}{c|}{$2^m$} & \multicolumn{1}{c|}{$m\geq 3; \nu\leq m; k\leq m-\nu;2|q$} & \multicolumn{1}{c|}{Optimal} \\
\hline
\multicolumn{1}{|c|}{Sarkar \cite{2019PalashZCCS}} & \multicolumn{1}{c|}{Boolean Functions} & \multicolumn{1}{c|}{$q$} & \multicolumn{1}{c|}{$2^{k+p+1}$} & \multicolumn{1}{c|}{$2^{k+1}$} & \multicolumn{1}{c|}{$2^{m-p}$} & \multicolumn{1}{c|}{$2^m$} & \multicolumn{1}{c|}{$m,k\geq 1;p\geq 0; 2|q$} & \multicolumn{1}{c|}{Optimal} \\
\hline
\multicolumn{1}{|c|}{\textit{Theorem  \ref{theorem:seed:Z:PU:unifying}}} & \multicolumn{1}{c|}{Block  Matrices} & \multicolumn{1}{c|}{$q$} & \multicolumn{1}{c|}{$K$} & \multicolumn{1}{c|}{$M$} & \multicolumn{1}{c|}{$M$} & \multicolumn{1}{c|}{$Kd^{N_0}_0d^{N_1}_1$} & \multicolumn{1}{c|}{$q\geq 2, d_0|K,d_1|M$; $N_0,N_1 \in \mathbb{N}$;  $K=MP$} & \multicolumn{1}{c|}{Not Optimal} \\
\hline
\multicolumn{1}{|c|}{\textit{Corollary \ref{corollary:seed:Z:PU}}} & \multicolumn{1}{c|}{$BH$ Matrices} & \multicolumn{1}{c|}{$q$} & \multicolumn{1}{c|}{$K$} & \multicolumn{1}{c|}{$M$} & \multicolumn{1}{c|}{$M$} & \multicolumn{1}{c|}{$K$} & \multicolumn{1}{c|}{$q\geq 2; M, K\geq 2$; $K=MP$} & \multicolumn{1}{c|}{Optimal} \\
\hline
\multicolumn{1}{|c|}{\textit{Theorem \ref{Proposed:Construction:Theorem}}} & \multicolumn{1}{c|}{ Length-$L$ ZPU Matrix} & \multicolumn{1}{c|}{$q$} & \multicolumn{1}{c|}{$K$} & \multicolumn{1}{c|}{$M$} & \multicolumn{1}{c|}{$MZ$} & \multicolumn{1}{c|}{$ML$} & \multicolumn{1}{c|}{$q\geq 2; M, K\geq 2$; $K=MP$} & \multicolumn{1}{c|}{Optimal} \\
\hline
\end{tabular}}
\end{center}

\section{\textbf{Preliminaries}} \label{Preli:sec:back}
In this section, we will present some basic definitions, notations and preliminaries. Also, we will provide a brief review of Butson-type Hadamard matrices.
\subsection{\textbf{ACCF and AACF}}
Given two complex-valued length-$L$ sequences $\textbf{\textit{x}}$ $=$ $\Big[x[0],$ $ x[1],$ $\cdots,$ $x[L-1]\Big]$ and $\textbf{\textit{y}}$ $=$ $\Big[y[0],$ $ y[1],$ $\cdots,$ $y[L-1]\Big]$, their aperiodic correlation function at time-shift $\tau$ is defined as
\begin{equation}
	\label{ACCF}
	R_{\textbf{\textit{x}}, \textbf{\textit{y}}}[\tau]= \begin{cases}
		\sum_{k=0}^{L-1-\tau}x[k]\cdot y^{*}[k+\tau] & 0\leq \tau \leq (L-1)\\
		\sum_{k=0}^{L-1+\tau}x[k-\tau]\cdot y^{*}[k] &-(L-1)\leq \tau < 0\\
		0; & \text{otherwise},
	\end{cases}
\end{equation} where $(\cdot)^*$ denotes complex conjugate. $R_{\textbf{\textit{x}}, \textbf{\textit{y}}}[\tau]$ is called aperiodic cross-correlation function (ACCF) when $\textbf{\textit{x}} \neq \textbf{\textit{y}}$; otherwise, it is called aperiodic auto-correlation function (AACF). For simplicity, AACF of $\textbf{\textit{x}}$ will be  written as $R_{\textbf{\textit{x}}}[\tau]$. Throughout this paper, a sequence is denoted by a bold Italian  lowercase letter. The $z$-transforms of the sequences $\textbf{\textit{x}}$ and $\textbf{\textit{y}}$ are defined by
\begin{align*}
	x(z)=\sum_{k=0}^{L-1}x[k]\cdot z^{-k} \qquad \text{and} \qquad y(z)=\sum_{k=0}^{L-1}y[k]\cdot z^{-k}.
\end{align*} We will use the convention $x^{*}(z)=\sum_{k=0}^{L-1}x^{*}[k]\cdot z^{-k}$. The sequence $x(z)$ is said to be a unimodular sequence if each coefficient of $x(z)$ has unit magnitude. According to $z$-transforms of $\textbf{\textit{x}}$ and $\textbf{\textit{y}}$, the $z$-transform of ACCF $R_{\textbf{\textit{x}}, \textbf{\textit{y}}}[\tau]$ is given by
\begin{equation}
R_{\textbf{\textit{x}}, \textbf{\textit{y}}}(z)=\sum_{\tau=-(L-1)}^{L-1}R_{\textbf{\textit{x}}, \textbf{\textit{y}}}[\tau]\cdot z^{-\tau}=x(z)\cdot y^{*}(z^{-1}).
\end{equation} For the given two sequence sets $\textbf{x}(z)=$ $\Big[ x_0(z),$  $x_1(z),$ $\cdots,$  $x_{M-1}(z)\Big]^T$\footnote{The sequence set $\textbf{x}(z)$ can also be denoted by $\textbf{x}=$ $\Big[ \textbf{\textit{x}}_0 ,$  $\textbf{\textit{x}}_1 ,$ $\cdots,$  $\textbf{\textit{x}}_{M-1}\Big]^T$ in time-domain, where $(\cdot)^T$ denotes the transpose operator.} and $\textbf{y}(z)=$ $\Big[ y_0(z),$   $y_1(z),$ $\cdots,$  $y_{M-1}(z)\Big]^T$ with equal length $L$, the ACCF sum $S_{\textbf{x},\textbf{y}}[\tau]$ between $\textbf{x}(z)$ and $\textbf{y}(z)$ at time-shift $\tau$ is defined by
\begin{equation}
S_{\textbf{x},\textbf{y}}[\tau]=\sum_{m=0}^{M-1}R_{\textbf{\textit{x}}_m,\textbf{\textit{y}}_m}[\tau].
\end{equation} The $z$-transform of ACCF sum  $S_{\textbf{x},\textbf{y}}[\tau]$ between $\textbf{x}(z)$ and $\textbf{y}(z)$ can be written as
\begin{align}
\label{ACCF_Sum:expression}
 S_{\textbf{x},\textbf{y}}(z)&=\sum_{\tau=-(L-1)}^{L-1}S_{\textbf{x},\textbf{y}}[\tau]\cdot z^{-\tau}=\sum_{\tau=-(L-1)}^{L-1}\sum_{m=0}^{M-1}R_{\textbf{\textit{x}}_m,\textbf{\textit{y}}_m}[\tau]\cdot z^{-\tau} \nonumber \\
 & ={\color{red}\sum_{\tau=-(L-1)}^{-Z}S_{\textbf{x},\textbf{y}}[\tau]\cdot z^{-\tau}+}{\color{blue}\sum_{\tau=-(Z-1)}^{Z-1}S_{\textbf{x},\textbf{y}}[\tau]\cdot z^{-\tau}}{\color{red}+\sum_{\tau=Z}^{L-1}S_{\textbf{x},\textbf{y}}[\tau]\cdot z^{-\tau}},
 \end{align} where $1\leq Z\leq L$. Note that we are focused on the aperiodic correlation sums between  the sets within the zone of length $Z$ throughout this paper. For this, we introduce a new function, called zone extraction function, for the desired correlation zone. This function will be extensively used later for the proof of our proposed ZCCSs. According to (\ref{ACCF_Sum:expression}), let us define the corresponding zone extraction function for the desired correlation zone as follows:
 \begin{definition}[Zone Extraction Function]
 \label{Zone:extraction:mapping}
 For given two sets $\textbf{x}(z)$ and $\textbf{y}(z)$ and $1\leq Z\leq L$, a zone extraction function $f_{Z}$ on $S_{\textbf{x},\textbf{y}}(z)$ is defined by
 \begin{equation}
 \label{equation:definition:zone:extraction}
 f_{Z}\Big(S_{\textbf{x},\textbf{y}}(z)\Big)={\color{blue}\sum_{\tau=-(Z-1)}^{Z-1}S_{\textbf{x},\textbf{y}}[\tau]\cdot z^{-\tau}}=\sum_{\tau=-(Z-1)}^{Z-1}\sum_{m=0}^{M-1}R_{\textbf{\textit{x}}_m,\textbf{\textit{y}}_m}[\tau]\cdot z^{-\tau}.
 \end{equation}
 \end{definition} The purpose of $f_{Z}$ is to collect all the correlation terms which have time-shifts less than $Z$. Clearly, $f_{Z}\Big(S_{\textbf{x},\textbf{y}}(z)\Big)=S_{\textbf{x},\textbf{y}}(z)$ when $Z=L$. We illustrate this function by the following example.
 \begin{example}
 Let $\textbf{x}=\bigl[\begin{smallmatrix}
 +++++\\
 +-+-+
\end{smallmatrix}\bigr]$ and $\textbf{y}=\bigl[\begin{smallmatrix}
 +++--\\
 +--+-
\end{smallmatrix}\bigr]$ be two sets of sequences with length $L=5$. Then, the $z$-transform of ACCF sum between $\textbf{x}$ and $\textbf{y}$ is given by
\begin{align}
S_{\textbf{x},\textbf{y}}(z)={\color{red}-2z^{4}+0z^{3}}{\color{blue}-4z^{2}+2z^{1}+0z^{0}
+2z^{-1}+4z^{-2}}{\color{red}+0z^{-3}+2z^{-4}}.
\end{align} For the desired correlation zone $-2$ to $2$, i.e., $Z=3$, a zone extraction function $f_{3}$ on $S_{\textbf{x},\textbf{y}}(z)$ is given by
\begin{align}
f_{3}\Big(S_{\textbf{x},\textbf{y}}(z)\Big)={\color{blue}-4z^{2}+2z^{1}+0z^{0}+2z^{-1}+4z^{-2}}.
\end{align} Note that the function $f_{3}$ takes all the correlation terms from $S_{\textbf{x},\textbf{y}}(z)$ within the time-shifts from $-2$ to $2$.
 \end{example}

 We remark that, throughout this paper, a ZCZ width is denoted by the upper case letter $Z$ (not to be confused with the indeterminate variable $z$ in $z$-transform). The uppercase and lowercase bold letters denote a matrix and a vector, respectively.

 \subsection{\textbf{Butson-type Hadamard (BH) Matrices}}
Butson-type Hadamard ($BH$) matrices play a very crucial role in the design of a large class of unimodular sequences with good correlation properties \cite{2017Shibsankar}, \cite{2018Shibsankar}. We provide a brief introduction here as we will use $BH$ matrices in our proposed constructions of ZPU matrices in Section \ref{Unifying:Framework:sec:pro}.

 A complex Hadamard matrix $\textbf{U}$ is an $M\times M$ complex matrix with unimodular entries such that $\textbf{U}^H\cdot \textbf{U}=M\textbf{I}_M$. A Butson-type Hadamard $BH(M,q)$ matrix refers to a complex Hadamard matrix of size $M\times M$ with $q^{\text{th}}$ roots of unity entries \cite{1962Butson}. That is, the elements of $BH(M,q)$ matrix are the powers of $q$-th root of unity. Note that the number of phases is $q$. It has been shown in \cite{1933Paley} that $BH(M,2)$ matrices exist only for $M=2,4m$, where $m$ is a positive integer.  A $BH(M,2)$ represents a binary Hadamard matrix, denoted by $\textbf{H}_{M}$ for $M=2,4m$ and $BH(M,M)$ represents discrete Fourier transform (DFT) matrix, denoted by $\textbf{F}_{M}$. Two $BH$ matrices with entries drawn from complex $q$-th roots of unity are said to be equivalent if one can be obtained from the other by a finite number of row permutations, column permutations, multiplication of a row by a complex $q$-th root of unity or multiplication of a column by a complex $q$-th root of unity. Any equivalence operation applied to a $BH$ matrix gives a $BH$ matrix. A matrix is said to be a normalised matrix if its first row and first column consist of 1s only. It follows that every $BH$ matrix is equivalent to a normalised $BH$ matrix. In \cite{2017Shibsankar}, it is shown that the use of equivalent forms of $BH$ matrices can significantly increase the number of complementary sequences.

In \cite{1962Butson}, Butson proved a necessary condition for the existence of a $BH(M,p)$ matrix, where $M=pt$ for a positive integer $t$ and a prime integer $p$. The problem of finding all the pairs $(M, q)$ such that $BH(M,q)$ matrix exist remains open. Moreover, the set of all $BH$ matrices is countable, but not finite. In \cite{2006Bruzda}, Bruzda \textit{et al.} have reported $BH$ matrices with size up to $M=16$. They also introduced methods to construct larger $BH$ matrices from smaller ones. In \cite{2008Compton}, Compton \textit{et al.} have shown that a $BH(M,6)$ does not exist if $M$ is odd and the squarefree part of $M$ is divisible by a prime $p\equiv 2$ (mod $3$). For instance, there is no $BH(M, 6)$ matrix when $M=5,11,15,17$, for $M\leq 19$. They have reported $BH(M,6)$ matrices for $M=2,3,4,6,7,8,9,10,12,13,14,16,18$. Later, Sz{\"o}ll{\H{o}}si proved that a $BH(19,6)$ matrix exists in \cite{2013Szollhosi}. According to the above discussion, we give the values of $M$ and $q$ for $BH(M,q)$ matrices up to $M=19$ in Table \ref{Table:BH:Matrices} \cite{2006Bruzda}\nocite{2008Compton}-\cite{2013Szollhosi}.
\begin{center}
\captionof{table}{The Values of $M$ and $q$ for $BH(M,q)$ Matrices up to $M=19$}  \label{Table:BH:Matrices}
	\resizebox{\linewidth}{!}{
		\renewcommand{\arraystretch}{1.8}
 \begin{tabular}{llllllllllllllllllllllllllllllllllllllllllllllllll}
\cline{1-49}
\multicolumn{1}{|c|}{$M$} & \multicolumn{2}{c|}{$2$} & \multicolumn{2}{c|}{$3$} & \multicolumn{3}{c|}{$4$} & \multicolumn{1}{c|}{$5$} & \multicolumn{3}{c|}{$6$} & \multicolumn{2}{c|}{$7$} & \multicolumn{4}{c|}{$8$} & \multicolumn{3}{c|}{$9$} & \multicolumn{4}{c|}{$10$} & \multicolumn{1}{c|}{$11$} & \multicolumn{5}{c|}{$12$} & \multicolumn{2}{c|}{$13$} & \multicolumn{5}{c|}{$14$} & \multicolumn{1}{c|}{$15$} & \multicolumn{5}{c|}{$16$} & \multicolumn{1}{c|}{$17$} & \multicolumn{2}{c|}{$18$} & \multicolumn{2}{c|}{$19$}   \\
\cline{1-49}
\multicolumn{1}{|c|}{$q$} & \multicolumn{1}{c|}{$2$} & \multicolumn{1}{c|}{$6$} & \multicolumn{1}{c|}{$3$} & \multicolumn{1}{c|}{$6$} & \multicolumn{1}{c|}{$2$} & \multicolumn{1}{c|}{$4$} & \multicolumn{1}{c|}{$6$} & \multicolumn{1}{c|}{$5$} & \multicolumn{1}{c|}{$3$} & \multicolumn{1}{c|}{$4$} & \multicolumn{1}{c|}{$6$} & \multicolumn{1}{c|}{$6$} & \multicolumn{1}{c|}{$7$} & \multicolumn{1}{c|}{$2$} & \multicolumn{1}{c|}{$4$} & \multicolumn{1}{c|}{$6$} & \multicolumn{1}{c|}{$8$} & \multicolumn{1}{c|}{$3$} & \multicolumn{1}{c|}{$6$} & \multicolumn{1}{c|}{$9$} & \multicolumn{1}{c|}{$4$} & \multicolumn{1}{c|}{$5$} & \multicolumn{1}{c|}{$6$} & \multicolumn{1}{c|}{$10$} & \multicolumn{1}{c|}{$11$} & \multicolumn{1}{c|}{$2$} & \multicolumn{1}{c|}{$3$} & \multicolumn{1}{c|}{$4$} & \multicolumn{1}{c|}{$6$} & \multicolumn{1}{c|}{$12$} & \multicolumn{1}{c|}{$6$} & \multicolumn{1}{c|}{$13$} & \multicolumn{1}{c|}{$4$} & \multicolumn{1}{c|}{$6$} & \multicolumn{1}{c|}{$7$} & \multicolumn{1}{c|}{$10$} & \multicolumn{1}{c|}{$14$} & \multicolumn{1}{c|}{$15$} & \multicolumn{1}{c|}{$2$} & \multicolumn{1}{c|}{$4$} & \multicolumn{1}{c|}{$6$} & \multicolumn{1}{c|}{$8$} & \multicolumn{1}{c|}{$16$} & \multicolumn{1}{c|}{$17$} & \multicolumn{1}{c|}{$6$} & \multicolumn{1}{c|}{$18$} & \multicolumn{1}{c|}{$6$} & \multicolumn{1}{c|}{$19$}   \\
\hline
\end{tabular}}
\end{center}
We give a $BH(6,3)$ matrix in the following example.
\begin{example}
 \label{BH(6,3):example}
 Let $M=6$ and $q=3$. Then, a $BH(6,3)$ matrix is given by
\begin{equation} \label{BH(6,3):spectral:matrix}
BH(6,3): \begin{bmatrix}
    0& 0 &0 &0& 0& 0   \\
    0 &0 &1  &1 & 2 &2  \\
    0 &1   &0& 2& 2&1 \\
    0&  1 &  2 & 0&  1 & 2\\
    0 &2&2& 1 & 0 & 1 \\
     0  &2& 1  & 2& 1  &0
\end{bmatrix},
\end{equation} where only the exponents of $\omega=e^{-2\pi \sqrt{-1} /3}$ are shown. This $BH(6,3)$ matrix was introduced by \cite{2003fuglede} as ``spectral matrix". Observe that all the entries of $BH(6,3)$ are powers of the cube root of unity $\omega$. That is, the elements of the $BH(6,3)$ matrix are drawn from a $3$-PSK constellation. In contrast, a $6\times 6$ DFT matrix $\textbf{F}_6$ lies upon a $6$-PSK constellation. We will use this $BH(6,3)$ matrix in Section \ref{sub:sec:optimal:ZPU}.
\end{example}
\subsection{\textbf{Matrices of Polynomials}}
In this paper, we use the term polynomial to mean $z$-transform of a sequence. A matrix of polynomials is simply a matrix whose entries are polynomials. Equivalently, it can be viewed as a polynomial with matrix coefficients.  We will use the following notation throughout this paper for the matrix of polynomials.

Let $\textbf{X}(z)= \Big[\textbf{x}_0(z), \textbf{x}_1(z),\cdots ,\textbf{x}_{K-1}(z)\Big]$ be a polynomial matrix of $K$ column vectors, each of size $M$, i.e.,
\begin{align}
\textbf{x}_\mu(z)=\Big[ x_{0\mu}(z),  x_{1\mu}(z), \cdots,  x_{(M-1)\mu}(z)\Big]^T,
\end{align} where $0 \leqslant \mu \leqslant K-1$ and $x_{m\mu}(z) $ is a polynomial over $z^{-1}$ having complex number coefficients and degree $L-1$ for each $m \in \{0,1,\cdots ,M-1\}$. The $z$-transform of ACCF sum $S_{\textbf{x}_\mu,\textbf{x}_\nu}[\tau]$ between two columns $\textbf{x}_\mu(z)$ and $\textbf{x}_\nu(z)$  ($0 \leq \mu, \nu \leq M-1$) is given by
\begin{equation}
	\label{ACCF:Sum:Condition}
S_{\textbf{x}_\mu,\textbf{x}_\nu}(z)=\sum_{m=0}^{M-1}R_{\textbf{\textit{x}}_{m\mu},\textbf{\textit{x}}_{m\nu}}(z)= \sum_{m=0}^{M-1}x_{m\mu}(z)\cdot x_{m\nu}^*(z^{-1})=\widetilde{\textbf{x}_\nu(z)}\cdot \textbf{x}_\mu(z),
\end{equation} where the tilde operator is defined by $\widetilde{\textbf{x}_\nu(z)}=\textbf{x}_\nu^{H}(z^{-1})$ and $(\cdot)^H$ is the Hermitian operation.  From $z$-transform of ACCF sum given by (\ref{ACCF:Sum:Condition}) and the tilde operation, the product $\widetilde{\textbf{X}(z)}\cdot\textbf{X}(z)$ of matrices can be expressed as
\begin{align}
\label{product:PU:condition:AACF:ACCF}
\widetilde{\textbf{X}(z)}\cdot\textbf{X}(z)&= \begin{bmatrix}
\widetilde{\textbf{x}_0(z)}\cdot \textbf{x}_0(z) & \widetilde{\textbf{x}_0(z)}\cdot \textbf{x}_1(z) & \cdots & \widetilde{\textbf{x}_0(z)}\cdot \textbf{x}_{K-1}(z)  \\
\widetilde{\textbf{x}_1(z)}\cdot \textbf{x}_0(z) & \widetilde{\textbf{x}_1(z)}\cdot \textbf{x}_1(z) & \cdots & \widetilde{\textbf{x}_1(z)}\cdot \textbf{x}_{K-1}(z) \\
\vdots & \vdots & \ddots & \vdots \\
\widetilde{\textbf{x}_{K-1}(z)}\cdot \textbf{x}_0(z) & \widetilde{\textbf{x}_{K-1}(z)}\cdot \textbf{x}_1(z) & \cdots & \widetilde{\textbf{x}_{K-1}(z)}\cdot \textbf{x}_{K-1}(z)
\end{bmatrix} \nonumber \\
&=\begin{bmatrix}
S_{\textbf{x}_0,\textbf{x}_0}(z) & S_{\textbf{x}_0,\textbf{x}_1}(z) & \cdots & S_{\textbf{x}_0,\textbf{x}_{K-1}}(z)  \\
S_{\textbf{x}_1,\textbf{x}_0}(z) & S_{\textbf{x}_1,\textbf{x}_1}(z) & \cdots & S_{\textbf{x}_1,\textbf{x}_{K-1}}(z)   \\
\vdots & \vdots & \ddots & \vdots \\
S_{\textbf{x}_{K-1},\textbf{x}_0}(z) & S_{\textbf{x}_{K-1},\textbf{x}_1}(z) & \cdots & S_{\textbf{x}_{K-1},\textbf{x}_{K-1}}(z)
\end{bmatrix}_{K\times K},
\end{align} where $\widetilde{\textbf{X}(z)}=\textbf{X}^{H}\left(z^{-1}\right)$ which is sometimes called the Hermitian version of $\textbf{X}(z^{-1})$. The matrix $\widetilde{\textbf{X}(z)}\cdot\textbf{X}(z)$ can be expressed only by the $z$-transforms of the AACF sums and ACCF sums between sequence sets (columns). That is, $\widetilde{\textbf{X}(z)}\cdot\textbf{X}(z)$ describes the matrix representation of ACCF sums between different columns of $\textbf{X}(z)$. We call $\widetilde{\textbf{X}(z)}\cdot\textbf{X}(z)$ the matrix of ACCF sums throughout this paper.

According to \textit{Definition \ref{Zone:extraction:mapping}} and (\ref{product:PU:condition:AACF:ACCF}), from now on, we will use the following convention:
\begin{equation}
f_{Z}\left(\widetilde{\textbf{X}(z)}\cdot\textbf{X}(z)\right)=\Big[f_{Z}\left(S_{\textbf{x}_\mu,\textbf{x}_\nu}(z)\right)\Big]_{K\times K},
\end{equation} where $f_{Z}$ is the zone extraction function defined by (\ref{equation:definition:zone:extraction}).
\subsection{\textbf{Paraunitary (PU) Matrix}}
A PU matrix is simply a matrix of polynomials over the indeterminate variable $z^{-1}$ which is unitary on the unit circle, i.e., $|z|=1$. That is, PU matrix is a generalization of unitary matrix.
\begin{definition}[\cite{1993Vaidyanathan}]
\label{Definition:PU}
An $M\times K$ polynomial matrix $\textbf{X}(z)$ over $z^{-1}$ is said to be a PU matrix if the following identity holds:
\begin{equation}\label{Definition:PU:ZCZ}
\widetilde{\textbf{X}(z)}\cdot\textbf{X}(z)=c\cdot\textbf{I}_{K},
\end{equation}
where $\textbf{I}_K$ is an  identity matrix  of size $K\times K$ and $c$ is a positive constant which gives the matrix energy.
\end{definition} Clearly, $f_{Z}\left(\widetilde{\textbf{X}(z)}\cdot\textbf{X}(z)\right)=\Big[f_{Z}\left(S_{\textbf{x}_\mu,\textbf{x}_\nu}(z)\right)\Big]_{K\times K}=\widetilde{\textbf{X}(z)}\cdot\textbf{X}(z)=c\cdot\textbf{I}_{K}$ when $Z=L$.
Equivalently, the above condition (\ref{Definition:PU:ZCZ}) can be written by
\begin{align}
\label{PU:Condition}
S_{\textbf{x}_\mu,\textbf{x}_\nu}(z)={\color{blue}c\cdot \delta(\mu-\nu)},
\end{align} where $\delta$ denotes the delta function. According to (\ref{product:PU:condition:AACF:ACCF}), (\ref{Definition:PU:ZCZ}) and (\ref{PU:Condition}), we can write the matrix of ACCF sums as follows:

\begin{align}
&\widetilde{\textbf{X}(z)}\cdot\textbf{X}(z)=\Big[S_{\textbf{x}_\mu,\textbf{x}_\nu}(z)\Big]_{K\times K} \nonumber \\
	&= \begin{bmatrix}
	{\color{blue}c+\sum_{1\leq |\tau|< L} 0\cdot z^{-\tau}} & {\color{blue}0+\sum_{1\leq |\tau|<L} 0\cdot z^{-\tau}} & \cdots &{\color{blue}0+\sum_{1\leq |\tau|<L} 0\cdot z^{-\tau}} \\
	{\color{blue}0+\sum_{1\leq |\tau|< L} 0\cdot z^{-\tau}} & {\color{blue}c+\sum_{1\leq |\tau|<L} 0\cdot z^{-\tau}} & \cdots & {\color{blue}0+\sum_{1\leq |\tau|<L} 0\cdot z^{-\tau}} \\
	\vdots & \vdots & \ddots & \vdots \\
	{\color{blue}0+\sum_{1\leq |\tau|<L} 0\cdot z^{-\tau}} & {\color{blue}0+\sum_{1\leq |\tau|<L} 0\cdot z^{-\tau}} & \cdots & {\color{blue}c+\sum_{1\leq |\tau|<L} 0\cdot z^{-\tau}} \\
\end{bmatrix}_{K\times K} \nonumber \\
&= \begin{bmatrix}
	{\color{blue}c} & {\color{blue}0} & \cdots &{\color{blue}0} \\
	{\color{blue}0} & {\color{blue}c} & \cdots & {\color{blue}0} \\
	\vdots & \vdots & \ddots & \vdots \\
	{\color{blue}0} & {\color{blue}0} & \cdots & {\color{blue}c} \\
\end{bmatrix}_{K\times K}= c\cdot\textbf{I}_{K}.
\end{align} Clearly, the matrix $\textbf{X}(z)$ satisfies the zero auto- and cross-correlation properties over the whole range of time-shifts from $-(L-1)$ to $(L-1)$ when it is a PU matrix. According to \cite{1993Vaidyanathan}, any arbitrary PU matrix can be factorized into a product of unitary and diagonal matrices. This factorization is said to be an expanded product form of a PU matrix. The degree of a PU matrix refers to the minimum number of delays required to implement it. The length of a PU matrix refers to the length of the constituent sequences. A PU matrix is called a unimodular PU matrix if it has only unimodular coefficients. For example, a PU matrix with $\pm 1$ coefficients refers to a binary PU matrix.

Based on the definitions of CCC and PU matrices, we state the following result on PU matrices.
\begin{result}[\cite{2018Shibsankar}] \label{Z-PU:result:equivalence:PU:CCC}
	The matrix $\textbf{X}(z)$ represents a polyphase $(M, M, L)$-CCC if and only if it is an $M\times M$ unimodular PU matrix of length $L$.
\end{result}
\begin{example} \label{CCC:Equi:PU}
Let $M=K=2$. A $2\times 2$ binary PU matrix $\textbf{X}(z)$ with sequence length $4$ is  given by
\begin{equation}
\textbf{X}(z)=\begin{bmatrix}
1+z^{-1}+z^{-2}-z^{-3} & 1+z^{-1}-z^{-2}+z^{-3} \\
1-z^{-1}+z^{-2}+z^{-3} & 1-z^{-1}-z^{-2}-z^{-3}
\end{bmatrix}_{2\times 2}.
\end{equation} It is easy to verify that $S_{\textbf{x}_{\mu}, \textbf{x}_{\nu}}(z)=c\cdot \delta(\mu -\nu)$, $\mu, \nu=0,1$. Therefore, we have the matrix of ACCF sums given by
\begin{align}
\widetilde{\textbf{X}(z)}\cdot\textbf{X}(z)&=\begin{bmatrix}
S_{\textbf{x}_0,\textbf{x}_0}(z) & S_{\textbf{x}_0,\textbf{x}_1}(z) \\
S_{\textbf{x}_1,\textbf{x}_0}(z) & S_{\textbf{x}_1,\textbf{x}_1}(z)
\end{bmatrix}_{2\times 2}  \nonumber \\
&=\begin{bmatrix}
{\color{blue}8+\sum_{1\leq |\tau|<4}0\cdot z^{-\tau}} & {\color{blue}0+\sum_{1\leq |\tau|<4}0\cdot z^{-\tau}} \\
{\color{blue}0+\sum_{1\leq |\tau|<4}0\cdot z^{-\tau}} & {\color{blue}8+\sum_{1\leq |\tau|<4}0\cdot z^{-\tau}}
\end{bmatrix}_{2\times 2} = \begin{bmatrix}
{\color{blue}8} & {\color{blue}0} \\
{\color{blue}0} & {\color{blue}8}
\end{bmatrix}_{2\times 2}= 8\cdot \textbf{I}_2.
\end{align} In this case, $f_{Z}\left(\widetilde{\textbf{X}(z)}\cdot\textbf{X}(z)\right)=\Big[f_{Z}\left(S_{\textbf{x}_\mu,\textbf{x}_\nu}(z)\right)\Big]_{2\times 2}=\Big[S_{\textbf{x}_\mu,\textbf{x}_\nu}(z)\Big]_{2\times 2}=8\cdot\textbf{I}_{2}$ for $Z=4$. That is, the ZCZ width of this PU matrix is over the whole range of time-shifts from $-3$ to $3$, i.e., $Z=L=4$.
\end{example}
Next, we recall our previous PU matrix construction for CCCs with flexible sequence lengths. We will use these PU matrices in the subsequent section.
\begin{lemma}[Construction of PU Matrices \cite{2018Shibsankar}]
	\label{SPL:2018}
	Let $M$ and $P$ be two positive integers which are greater than one such that $P|M$. We consider $BH$ matrices $\textbf{A}_{n}$ and $\textbf{U}_{0}$ of size $P\times P$ and $M\times M$, respectively. We first take the following two matrices
	\begin{align}
	\bm{\mathcal{U}}_{n}=& \textbf{I}_{\frac{M}{P}}\otimes \textbf{A}_{n}, \label{Kronecker:unitary:PXP}\\
	\bm{\mathcal{D}}(z)=& \textbf{I}_{\frac{M}{P}}\otimes \text{diag}\Big(1,z^{-1},\cdots, z^{-(P-1)}\Big), \label{Kronecker:Diag:PXP}
	\end{align} where $n\in \{1,2,\cdots ,N\}$ for each positive integer $N$ and $\otimes$ is Kronecker product. Then, our recursive generator for an $M\times M$ PU matrix is given by
	\begin{align} \label{SPL:new:algorithm}
	\bm{\mathcal{G}}_{n}(z)=\bm{\mathcal{U}}_{n}\cdot \bm{\mathcal{D}}\left(z^{D_n}\right)\cdot \textbf{P}_{n} \cdot \bm{\mathcal{G}}_{n-1}(z) \cdot \textbf{Q}_{n},
	\end{align} where $\bm{\mathcal{G}}_{0}(z)=\textbf{U}_{0}$, $D_n=P^{\pi(n-1)}$, $\pi$ is an arbitrary permutation of the numbers $\{0,1,\cdots,N-1\},$ and $\textbf{P}_{n}, \textbf{Q}_{n}$ are two arbitrary permutation matrices of equal size $M\times M$. Then, $\bm{\mathcal{G}}_{N}(z)$ is an $M\times M$ PU matrix of sequence length $P^N$.
\end{lemma} We give the example below to illustrate \textit{Lemma \ref{SPL:2018}}. In this example, we consider a small sequence length as we will use the constructed matrix for ZCCS construction in Section \ref{subsection:unifying:construction}.
\begin{example}
	\label{example:SPL:2018:Flexible}
	Let $M=6, P=2$, $N=1,$ and $\pi=[0]$. Also, let $\textbf{P}_{n}=\textbf{Q}_{n}=\textbf{I}_{6}$, $\textbf{A}_{n}=\textbf{H}_{2},$ and $\textbf{U}_{0}=\textbf{S}_{6}$ with $3$-PSK constellation given by (\ref{BH(6,3):spectral:matrix}) in \textit{Example \ref{BH(6,3):example}}. We have $D_{1}=P^{\pi(0)}=1$. Applying (\ref{Kronecker:unitary:PXP}) and (\ref{Kronecker:Diag:PXP}), we have $\bm{\mathcal{U}}_{n}=\textbf{I}_{3}\otimes \textbf{H}_{2}$ and  $\bm{\mathcal{D}}(z)= \textbf{I}_{3}\otimes \text{diag}\left(1, z^{-1}\right)$. Then, a $6\times 6$ PU matrix with sequence length $P^N=2$ can be written by
	\begin{align} \label{generating:matrix:12x12}
	&\bm{\mathcal{G}}_{1}(z)=\bm{\mathcal{U}}_{1}\cdot \bm{\mathcal{D}}\left(z^{D_1}\right)\cdot \textbf{U}_0=\Big(\textbf{I}_{3}\otimes \textbf{H}_{2}\cdot \text{diag}\left(1, z^{-1}\right)\Big)\cdot \textbf{S}_{6}.
	\end{align} Based on \textit{Result \ref{Z-PU:result:equivalence:PU:CCC}}, we call it as $(6,6,2)$-CCC. The CCC equivalent to this $6\times 6$ PU matrix is given in Table \ref{Table:PU:Matrix:L2}, where only exponents of $\omega=e^{-2\pi \sqrt{-1}/6}$ are given. Note that the number of phases of the constructed sequences is $LCM(2,3)=6$, where $LCM$ denotes the least common multiple.
	\begin{center}
\captionof{table}{A $6 \times 6$ PU Matrix with Sequence Length $2$}  \label{Table:PU:Matrix:L2}
	%\resizebox{\linewidth}{!}{
		\renewcommand{\arraystretch}{0.9}
	\begin{tabular}{|l|l|l|l|l|l|l|l|l|l|l|l|}
\hline
\multicolumn{1}{|c|}{} & \multicolumn{1}{c|}{$0 0$} & \multicolumn{1}{c|}{} & \multicolumn{1}{c|}{$0 0$} & \multicolumn{1}{c|}{} & \multicolumn{1}{c|}{$0 2$} & \multicolumn{1}{c|}{} & \multicolumn{1}{c|}{$0 2$} & \multicolumn{1}{c|}{} & \multicolumn{1}{c|}{$0 4$} & \multicolumn{1}{c|}{} & \multicolumn{1}{c|}{$0 4$} \\
\multicolumn{1}{|c|}{} & \multicolumn{1}{c|}{$0 3$} & \multicolumn{1}{c|}{} & \multicolumn{1}{c|}{$0 3$} & \multicolumn{1}{c|}{} & \multicolumn{1}{c|}{$0 5$} & \multicolumn{1}{c|}{} & \multicolumn{1}{c|}{$0 5$} & \multicolumn{1}{c|}{} & \multicolumn{1}{c|}{$0 1$} & \multicolumn{1}{c|}{} & \multicolumn{1}{c|}{$0 1$} \\
\multicolumn{1}{|c|}{$\textbf{g}_0$} & \multicolumn{1}{c|}{$0 0$} & \multicolumn{1}{c|}{$\textbf{g}_1$} & \multicolumn{1}{c|}{$2 2$} & \multicolumn{1}{c|}{$\textbf{g}_2$} & \multicolumn{1}{c|}{$0 4$} & \multicolumn{1}{c|}{$\textbf{g}_3$} & \multicolumn{1}{c|}{$4 0$} & \multicolumn{1}{c|}{$\textbf{g}_4$} & \multicolumn{1}{c|}{$4 2$} & \multicolumn{1}{c|}{$\textbf{g}_5$} & \multicolumn{1}{c|}{$2 4$} \\
\multicolumn{1}{|c|}{} & \multicolumn{1}{c|}{$0 3$} & \multicolumn{1}{c|}{} & \multicolumn{1}{c|}{$2 5$} & \multicolumn{1}{c|}{} & \multicolumn{1}{c|}{$0 1$} & \multicolumn{1}{c|}{} & \multicolumn{1}{c|}{$4 3$} & \multicolumn{1}{c|}{} & \multicolumn{1}{c|}{$4 5$} & \multicolumn{1}{c|}{} & \multicolumn{1}{c|}{$2 1$} \\
\multicolumn{1}{|c|}{} & \multicolumn{1}{c|}{$0 0$} & \multicolumn{1}{c|}{} & \multicolumn{1}{c|}{$4 4$} & \multicolumn{1}{c|}{} & \multicolumn{1}{c|}{$4 2$} & \multicolumn{1}{c|}{} & \multicolumn{1}{c|}{$2 4$} & \multicolumn{1}{c|}{} & \multicolumn{1}{c|}{$0 2$} & \multicolumn{1}{c|}{} & \multicolumn{1}{c|}{$2 0$} \\
\multicolumn{1}{|c|}{} & \multicolumn{1}{c|}{$0 3$} & \multicolumn{1}{c|}{} & \multicolumn{1}{c|}{$4 1$} & \multicolumn{1}{c|}{} & \multicolumn{1}{c|}{$4 5$} & \multicolumn{1}{c|}{} & \multicolumn{1}{c|}{$2 1$} & \multicolumn{1}{c|}{} & \multicolumn{1}{c|}{$0 5$} & \multicolumn{1}{c|}{} & \multicolumn{1}{c|}{$2 3$} \\
\hline
\end{tabular}
\end{center}
\end{example}

\subsection{\textbf{$Z$-Complementary Code Sets (ZCCS)}}
A set $\textbf{x}(z)$ of $M$ unimodular sequences with equal length $L$ is called a $Z$-complementary code (ZCC) \cite{2007Fan_ZCSS} if
\begin{equation}
S_{\textbf{x}}(z)=\sum_{m=0}^{M-1}R_{\textbf{\textit{x}}_m}(z)={\color{blue}ML}{\color{red}+\sum_{\tau=Z}^{L-1} S_{\textbf{x}}[\tau]\cdot z^{-\tau}},
\end{equation} where $Z$ is called zero correlation zone (ZCZ). When $Z=L$, a ZCC reduces to a conventional complementary set of sequences. By an $(M, L, Z)$-ZCC, we mean a ZCC of $M$ sequences with length $L$ and ZCZ width $Z$.

A ZCC $\textbf{y}(z)$ of $M$ unimodular sequences with length $L$ is said to be a $Z$-complementary mate of ZCC $\textbf{x}(z)$ if
\begin{equation}
S_{\textbf{x},\textbf{y}}(z)=\sum_{m=0}^{M-1}R_{\textbf{\textit{x}}_m,\textbf{\textit{y}}_m}(z)={\color{blue}0}{\color{red}+\sum_{\tau=Z}^{L-1} S_{\textbf{x},\textbf{y}}[\tau]\cdot z^{-\tau}}.
\end{equation} Clearly, a $Z$-complementary mate becomes a conventional complementary mate when $Z=L$.  For a given complementary set of size $M$, it has been shown that there exist at most $M$ distinct complementary mates. However, for a given ZCC of size $M$, there exist more than $M$ distinct $Z$-complementary mates.

\begin{definition}
	The family $\textbf{X}(z)$ is called a $Z$-complementary code set (ZCCS) if each set is ZCC and two distinct sets are $Z$-complementary mates.
\end{definition} We denote it as $(K,Z)$-ZCCS$^L_{M}$. Obviously, $(K,Z)$-ZCCS$^L_{M}$ becomes a conventional mutually orthogonal complementary sets of sequences (MOCSS)\footnote{A family of sequence sets refers to a mutually orthogonal complementary sets of sequences (MOCSS) if the AACF sum of each set is zero except for zero time-shift and the ACCF sum between two distinct sets is zero for
any time-shifts \cite{1988NSuehiro}. When the set size achieves the upper bound, a MOCSS becomes a set of CCC.} when $Z=L$. In fact, a $(K,Z)$-ZCCS$^L_{M}$ becomes a set of CCC when $Z=L$ and $K=M$. For any given $(K,Z)$-ZCCS$^L_{M}$, the theoretical set size upper bound \cite{2007Fan_ZCSS}, \cite{2011LiuGuanNgChen} is given by
\begin{equation}
\label{mathematical:upper:bound:K}
K\leq M \lfloor{L/Z}\rfloor,
\end{equation} where $\lfloor{a}\rfloor$ represents the largest integer smaller than or equal to $a$.

We are now ready to introduce the concept of $Z$-paraunitary (ZPU) matrices. Later, we will propose a novel construction framework for optimal ZPU matrices with regard to the set size upper bound.
\section{\textbf{Concept of $Z$-Paraunitary Matrices}}
\label{Z-PU:Sec:Introduction}
In this section, we will first introduce the concept of ZPU matrices. Then, we will study some interesting properties of ZPU matrices. Finally, we will show the relationship between ZPU matrix and $(K,Z)$-ZCCS$^L_{M}$.

\subsection{\textbf{$Z$-Paraunitary (ZPU) Matrices}}
\begin{figure}
	\centering
	$\begin{bmatrix}
\includegraphics[width=0.27\textwidth]{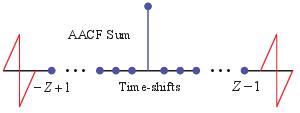} &
\includegraphics[width=0.27\textwidth]{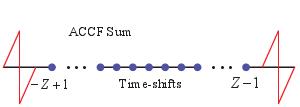} & \cdots & \includegraphics[width=0.27\textwidth]{Z_PU_ACCF.jpg} \\
\includegraphics[width=0.27\textwidth]{Z_PU_ACCF.jpg} &
\includegraphics[width=0.27\textwidth]{Z_PU_AACF.jpg} & \cdots & \includegraphics[width=0.27\textwidth]{Z_PU_ACCF.jpg} \\
	\vdots & \vdots & \ddots & \vdots \\
	\includegraphics[width=0.27\textwidth]{Z_PU_ACCF.jpg} & \includegraphics[width=0.27\textwidth]{Z_PU_ACCF.jpg} & \cdots & \includegraphics[width=0.27\textwidth]{Z_PU_AACF.jpg}
	\end{bmatrix}_{K\times K}$
	\caption{The time-domain expression of $\widetilde{\textbf{X}(z)}\cdot\textbf{X}(z)$ for a ZPU matrix $\textbf{X}(z)$.}
	\label{Figure:Z-PU:time:domain}
\end{figure}

\begin{definition}
\label{definition:Z-PU:matrix}
	An $M\times K$ matrix $\textbf{X}(z)$ of polynomials over $z^{-1}$ is said to be a ZPU matrix if the following identity holds:
	\begin{equation}
	 \label{Definition:ZPU:Condition}
	 f_{Z}\Big(\widetilde{\textbf{X}(z)}\cdot\textbf{X}(z)\Big)=c\cdot\textbf{I}_{K},
	 \end{equation} where $f_{Z}$ is the zone extraction function defined by (\ref{equation:definition:zone:extraction}) for the desired ZCZ width $Z$.
\end{definition}  It is worth noting that we are focused on the aperiodic correlation sums between the sets within the zone of length $Z$. Therefore, only the correlation terms from time-shifts $-(Z-1)$ to $(Z-1)$ of $S_{\textbf{x}_\mu,\textbf{x}_\nu}(z)$ are taken into consideration. According to (\ref{product:PU:condition:AACF:ACCF}), an equivalent expression of (\ref{Definition:ZPU:Condition}) can be written as
\begin{align}
\label{Z_PU_ACCFS_SUm_condition}
S_{\textbf{x}_\mu,\textbf{x}_\nu}(z)={\color{red}\sum_{\tau=-(L-1)}^{-Z} S_{\textbf{x}_\mu,\textbf{x}_\nu}[\tau]\cdot z^{-\tau}+}{\color{blue}c\cdot \delta(\mu-\nu)}{\color{red}+\sum_{\tau=Z}^{L-1} S_{\textbf{x}_\mu,\textbf{x}_\nu}[\tau]\cdot z^{-\tau}}, \ 0\leq \mu, \nu \leq K-1.
\end{align} Based on (\ref{equation:definition:zone:extraction}), we have $f_{Z}\Big(S_{\textbf{x}_\mu,\textbf{x}_\nu}(z)\Big)={\color{blue}c\cdot \delta(\mu-\nu)}$ for the ZCZ width $Z$.  We call this matrix a ZPU matrix of size $M\times K$ and sequence length $L$ (i.e., degree $L-1$).
According to (\ref{product:PU:condition:AACF:ACCF}), (\ref{Definition:ZPU:Condition}) and (\ref{Z_PU_ACCFS_SUm_condition}), we can write explicitly the matrix of ACCF sums as follows:
	\begin{align}
	f_{Z}\left(\widetilde{\textbf{X}(z)}\cdot\textbf{X}(z)\right)&=\Big[f_{Z}\left(S_{\textbf{x}_\mu,\textbf{x}_\nu}(z)\right)\Big]_{K\times K}  \nonumber \\
	&=  \begin{bmatrix}
	f_{Z}\left({\color{blue}c\cdot \delta(\mu-\nu)}{\color{red}+\sum_{Z\leq |\tau|<L} S_{\textbf{x}_\mu,\textbf{x}_\nu}[\tau]\cdot z^{-\tau}}\right)
	\end{bmatrix}_{K\times K} \nonumber \\
	&= \begin{bmatrix}
	{\color{blue}c\cdot \delta(\mu-\nu)}
	\end{bmatrix}_{K\times K} \nonumber \\
&= \begin{bmatrix}
	{\color{blue}c} & {\color{blue}0} & \cdots & {\color{blue}0} \\
	{\color{blue}0}& {\color{blue}c} & \cdots  &{\color{blue}0}\\
	\vdots & \vdots &\ddots & \vdots \\
	{\color{blue}0} & {\color{blue}0} & \cdots & {\color{blue}c}
\end{bmatrix}_{K\times K} \nonumber \\
&= c\cdot\textbf{I}_{K}.
	\end{align} An equivalent time-domain expression of $\widetilde{\textbf{X}(z)}\cdot\textbf{X}(z)$ for a ZPU matrix is given in Fig. \ref{Figure:Z-PU:time:domain}.  One can notice that we have relaxed the condition on the range of time-shifts $Z\leq L$ in (\ref{Definition:ZPU:Condition}). Later, we will show that the number of column vectors (sets) can be much larger than the size of the column vector (i.e., the number of constituent sequences). In other words, $K$ can be much larger than $M$. Clearly, \textit{Definition \ref{definition:Z-PU:matrix}} includes \textit{Definition \ref{Definition:PU}} as a special case when $Z=L$. Similar to a PU matrix, we can define the degree and length for a ZPU matrix.

 \subsection{\textbf{Properties of ZPU matrices}} In this subsection, we will investigate some important properties of ZPU matrices. We will use these properties in Section \ref{Unifying:Framework:sec:pro}.
\begin{property}
For an $M\times K$ ZPU matrix $\textbf{X}(z)$, the conjugate matrix $\textbf{X}^{*}(z)$ is also a ZPU matrix.
\end{property}
\begin{property}
\label{property:length:extension}
Let $\textbf{X}(z)$ be a PU matrix and $\textbf{Y}(z)$ be a ZPU matrix with size $M\times K$ and $K\times N$, respectively. Then, their product matrix $\textbf{X}(z)\cdot \textbf{Y}(z)$ is also a ZPU matrix with size $M\times N$.
\end{property}
\begin{IEEEproof}
Let $\textbf{W}(z)=\textbf{X}(z)\cdot \textbf{Y}(z)$. Clearly, $\textbf{W}(z)$ is a polynomial matrix of size $M\times N$.
Since $\textbf{X}(z)$ is a PU matrix and $\textbf{Y}(z)$ is a ZPU matrix, we have
\begin{align*}
&\widetilde{\textbf{X}(z)}\cdot \textbf{X}(z)= c\cdot \textbf{I}_{K}, \ \text{and} \ f_{Z}\left(\widetilde{\textbf{Y}(z)}\cdot \textbf{Y}(z)\right)= c'\cdot \textbf{I}_{N}.
\end{align*} Then, we have
\begin{align}
f_{Z}\left(\widetilde{\textbf{W}(z)}\cdot \textbf{W}(z)\right)&=f_{Z}\left(\widetilde{\textbf{X}(z)\cdot \textbf{Y}(z)} \cdot \textbf{X}(z)\cdot \textbf{Y}(z)\right) \nonumber \\
&=f_{Z}\left(\widetilde{\textbf{Y}(z)} \cdot \widetilde{\textbf{X}(z)} \cdot \textbf{X}(z)\cdot \textbf{Y}(z)\right) \nonumber \\
&= f_{Z}\left(\widetilde{\textbf{Y}(z)} \cdot c\textbf{I}_K\cdot \textbf{Y}(z)\right) \quad (\because \text{$\textbf{X}(z)$ is a PU matrix}) \nonumber \\
&= c\cdot f_{Z}\left(\widetilde{\textbf{Y}(z)} \cdot \textbf{Y}(z)\right) \nonumber \\
&= cc' \cdot \textbf{I}_N  \quad (\because \text{$\textbf{Y}(z)$ is a ZPU matrix})
\end{align} This completes the proof.
\end{IEEEproof}
\begin{remark} Note that the ZCZ width remains the same in \textit{Property \ref{property:length:extension}}, but it will allow to extend the sequence length of a given ZPU matrix. Later, we will show that both sequence length and ZCZ width can also be extended proportionally by meeting the set size upper bound.
\end{remark}

For the given two matrices of polynomials $\textbf{X}(z)$ and $\textbf{Y}(z)$ with size $M_1\times K_1$, and $M_2\times K_2$, respectively, their Kronecker product $\textbf{X}(z)\otimes \textbf{Y}(z)$ is also a matrix of polynomials with size $M_1M_2\times K_1K_2$ \cite{2005Phoong}. According to the tilde operation, we have
\begin{equation} \label{tilde{Kronecker Product}}
\widetilde{\textbf{X}(z)\otimes \textbf{Y}(z)}=\widetilde{\textbf{X}(z)} \otimes \widetilde{\textbf{Y}(z)},
\end{equation} where $\otimes$ denotes Kronecker product.
Let $\textbf{X}(z), \textbf{Y}(z), \textbf{Z}(z),$ and $\textbf{T}(z)$ be four matrices of polynomials with size $M\times K, N\times P,K\times R$, and $P\times S$, respectively. Then, the product rule is given by
\begin{align} \label{tilde{Product Rule}}
\Big(\textbf{X}(z)\otimes \textbf{Y}(z)\Big)& \cdot \Big(\textbf{Z}(z)\otimes \textbf{T}(z)\Big)= \Big(\textbf{X}(z)\cdot \textbf{Z}(z)\Big)\otimes \Big(\textbf{Y}(z)\cdot \textbf{T}(z)\Big).
\end{align} Based on the above properties of Kronecker product, we have the following method to enlarge the size of a ZPU matrix.
\begin{property}
\label{set:size:enlarge:property}
Let $\textbf{X}(z)$ be a ZPU matrix and $\textbf{Y}(z)$ be a PU matrix with size $M_1\times K_1$ and $M_2\times K_2$, respectively. Then, the matrix $\textbf{X}(z)\otimes \textbf{Y}(z)$ is also a ZPU matrix with size $M_1M_2\times K_1K_2$.
\end{property}
\begin{IEEEproof}
It is clear that $\textbf{X}(z)\otimes \textbf{Y}(z)$ is a polynomial matrix of size $M_1M_2\times K_1K_2$. Since $\textbf{X}(z)$ is a ZPU matrix and $\textbf{Y}(z)$ is a PU matrix, we have
\begin{align*}
&f_{Z}\left(\widetilde{\textbf{X}(z)}\cdot \textbf{X}(z)\right)= c_1\cdot \textbf{I}_{K_1}\ \text{and}\ \widetilde{\textbf{Y}(z)}\cdot \textbf{Y}(z)=c_2\cdot \textbf{I}_{K_2}.
\end{align*} By using the properties of Kronecker product, we can write
\begin{align}
f_{Z}\left(\widetilde{\textbf{X}(z)\otimes \textbf{Y}(z)} \cdot \textbf{X}(z)\otimes \textbf{Y}(z)\right)&=f_{Z}\left(\widetilde{\textbf{X}(z)}\otimes \widetilde{\textbf{Y}(z)} \cdot \textbf{X}(z)\otimes \textbf{Y}(z)\right) \quad \Big(\text{Using (\ref{tilde{Kronecker Product}})} \Big) \nonumber \\
&=f_{Z}\left(\left(\widetilde{\textbf{X}(z)}\cdot \textbf{X}(z)\right) \otimes \left(\widetilde{\textbf{Y}(z)}\cdot \textbf{Y}(z)\right)\right) \quad \Big(\text{Using (\ref{tilde{Product Rule}})} \Big) \nonumber \\
&=f_{Z}\left(\widetilde{\textbf{X}(z)}\cdot \textbf{X}(z)\right) \otimes \left(\widetilde{\textbf{Y}(z)}\cdot \textbf{Y}(z)\right) \nonumber \\
&= c_1\cdot \textbf{I}_{K_1} \otimes c_2\cdot \textbf{I}_{K_2} \nonumber \\
&= c_1c_2\cdot \textbf{I}_{K_1K_2}.
\end{align} This completes the proof.
\end{IEEEproof}
\begin{remark}
One can observe that \textit{Property \ref{set:size:enlarge:property}} allows only to enlarge the size of a given ZPU matrix by keeping the same ZCZ width.
\end{remark}

 \subsection{\textbf{Relationship between ZPU matrix  and ZCCS}}
  From the definitions of ZPU matrix  and ZCCS, we have the following property.
\begin{property}
\label{property:Z-CSS:Z-PU}
A polynomial matrix $\textbf{X}(z)$ is a ZPU matrix with size $M\times K$ and sequence length $L$ if and only if it is a $(K,Z)$-ZCCS$^L_{M}$.
\end{property}
\begin{IEEEproof}
For any two columns $\textbf{x}_\mu(z)$ and $\textbf{x}_\nu(z)$ of $\textbf{X}(z)$, we can write
\begin{equation}
S_{\textbf{x}_\mu,\textbf{x}_\nu}(z)=\sum_{m=0}^{M-1}R_{\textbf{\textit{x}}_{m\mu},\textbf{\textit{x}}_{m\nu}}(z)= \widetilde{\textbf{x}_\nu(z)}\cdot \textbf{x}_\mu(z),
\end{equation} where $\textbf{x}_\mu(z)=\Big[ x_{0\mu}(z),  x_{1\mu}(z), \cdots,  x_{(M-1)\mu}(z)\Big]^T$ consisting of $M$ unimodular sequences of equal length $L$ and $0\leq \mu, \nu\leq K-1$. Then, we have
{\small
\begin{align*}
\label{ZPU:Condition:Theorem}
&S_{\textbf{x}_\mu,\textbf{x}_\nu}(z)={\color{red}\sum_{\tau=-(L-1)}^{-Z} S_{\textbf{x}_\mu,\textbf{x}_\nu}[\tau]\cdot z^{-\tau}}+{\color{blue}c\cdot \delta(\mu-\nu)}+{\color{red}\sum_{\tau=Z}^{L-1} S_{\textbf{x}_\mu,\textbf{x}_\nu}[\tau]\cdot z^{-\tau}}\ \text{if and only if} \ f_{Z}\left(\widetilde{\textbf{X}(z)}\cdot\textbf{X}(z)\right)= c\cdot\textbf{I}_{K}.
\end{align*}} Thus, the matrix $\textbf{X}(z)$ is a ZPU matrix  $\Leftrightarrow f_{Z}\left(\widetilde{\textbf{X}(z)}\cdot\textbf{X}(z)\right)= c\cdot\textbf{I}_{K}\Leftrightarrow S_{\textbf{x}_\mu,\textbf{x}_\nu}(z)={\color{red}\sum_{\tau=-(L-1)}^{-Z} S_{\textbf{x}_\mu,\textbf{x}_\nu}[\tau]\cdot z^{-\tau}}$ $+{\color{blue}c\cdot \delta(\mu-\nu)}+{\color{red}\sum_{\tau=Z}^{L-1} S_{\textbf{x}_\mu,\textbf{x}_\nu}[\tau]\cdot z^{-\tau}}\Leftrightarrow\textbf{X}(z)$ is a $(K,Z)$-ZCCS$^L_{M}$. This completes the proof.
\end{IEEEproof}
\begin{remark}
According to  \textit{Property \ref{property:Z-CSS:Z-PU}}, there exists an equivalence between ZCCS and ZPU matrix. This equivalence allows us to find more binary ZPU matrices with odd length $\geq 3$.
\end{remark}

Note that when ZCCS is expressed as a ZPU matrix, each column corresponds to a ZCC and each entry (as a polynomial over $Z^{-1}$) of such a column corresponds to a constituent sequence of this ZCC. Also, two distinct columns correspond to $Z$-complementary mates. According to (\ref{mathematical:upper:bound:K}), a ZPU matrix of size $M\times K$ and length $L$ is said to be an optimal ZPU matrix if $K=M\lfloor{L/Z}\rfloor$. In general, $K$ can be much larger than $M$ for an $M\times K$ ZPU matrix. For an $M\times K$ PU matrix, the mathematical upper bound of $K$ follows from the special case when $Z=L$.

\begin{remark} We observe that \textit{Property \ref{property:Z-CSS:Z-PU}} includes \cite[Th. 1]{2018Shibsankar} as a special case when $M=K$ and $Z=L$.
\end{remark}

\subsection{\textbf{Examples of ZPU matrices}}
In this subsection, we give examples of both square and non-square ZPU matrices to illustrate our proposed concept.
\begin{example}
\label{example:2_2_3_2_PU}
Let $M=K=2$ and $L=3$. A binary polynomial matrix $\textbf{X}(z)$ of size $2\times 2$ and length $L=3$ is given by
\begin{equation}
\textbf{X}(Z)=\begin{bmatrix}
1+z^{-1}+z^{-2} & 1-z^{-1}+z^{-2} \\
1-z^{-1}+z^{-2} & -1-z^{-1}-z^{-2}
\end{bmatrix}_{2\times 2}.
\end{equation} Then, the matrix of ACCF sums is given by
\begin{align}
&\widetilde{\textbf{X}(z)}\cdot\textbf{X}(z)=\begin{bmatrix}
S_{\textbf{x}_0,\textbf{x}_0}(z) & S_{\textbf{x}_0,\textbf{x}_1}(z) \\
S_{\textbf{x}_1,\textbf{x}_0}(z) & S_{\textbf{x}_1,\textbf{x}_1}(z)
\end{bmatrix}_{2\times 2}  \nonumber \\
&=\begin{bmatrix}
{\color{red}2z^{2}+}{\color{blue}0z^{1}+6+0z^{-1}}{\color{red}+2z^{-2}} & {\color{red}0z^{2}+}{\color{blue}0z^{1}+0+0z^{-1}}{\color{red}+0z^{-2}} \\
{\color{red}0z^{2}+}{\color{blue}0z^1+0+0z^{-1}}{\color{red}+0z^{-2}} & {\color{red}2z^{2}+}{\color{blue}0z^1+6+0z^{-1}}{\color{red}+2z^{-2}}
\end{bmatrix}_{2\times 2}.
\end{align} By applying the zone extraction function for the time-shifts from $-1$ to $1$, i.e., $Z=2$, we have
\begin{align}
f_{Z}\left(\widetilde{\textbf{X}(z)}\cdot\textbf{X}(z)\right)&=\left[f_{Z}\left(S_{\textbf{x}_\mu,\textbf{x}_\nu}(z)\right)\right]_{2\times 2} \nonumber \\
&=\begin{bmatrix}
{\color{blue}0z^{1}+6+0z^{-1}} & {\color{blue}0z^{1}+0+0z^{-1}} \\
{\color{blue}0z^1+0+0z^{-1}} & {\color{blue}0z^1+6+0z^{-1}}
\end{bmatrix}_{2\times 2} \nonumber \\
&=\begin{bmatrix}
{\color{blue}6} & {\color{blue}0} \\
{\color{blue}0} & {\color{blue}6}
\end{bmatrix}_{2\times 2} =6\cdot \textbf{I}_2
\end{align} for the ZCZ width $Z=2$  and hence $\textbf{X}(z)$ is a binary $2$-PU matrix, i.e., it is a binary PU matrix within the ZCZ width $2$. Here, the function $f_{Z}$ picks the correlation terms from $S_{\textbf{x}_\mu,\textbf{x}_\nu}(z)$ within the correlation zone $-(Z-1)$ to $(Z-1)$.
\end{example}
\begin{example}
\label{example:2_2_5_2_PU}
Let $M=K=2$ and $L=5$. A binary polynomial matrix $\textbf{X}(z)$ of size $2\times 2$ and length $L=5$ is given by
\begin{equation}
\textbf{X}(z)=\begin{bmatrix}
1+z^{-1}+z^{-2}+z^{-3}-z^{-4} & -1+z^{-1}+z^{-2}-z^{-3}+z^{-4} \\
1-z^{-1}+z^{-2}+z^{-3}-z^{-4} & 1-z^{-1}-z^{-2}-z^{-3}-z^{-4}
\end{bmatrix}_{2\times 2}.
\end{equation} Then, the matrix of ACCF sums is given by
\begin{align}
&\widetilde{\textbf{X}(z)}\cdot\textbf{X}(z)=\begin{bmatrix}
S_{\textbf{x}_0,\textbf{x}_0}(z) & S_{\textbf{x}_0,\textbf{x}_1}(z) \\
S_{\textbf{x}_1,\textbf{x}_0}(z) & S_{\textbf{x}_1,\textbf{x}_1}(z)
\end{bmatrix}_{2\times 2}  \nonumber \\
&= \left[\begin{matrix}
{\color{red}-2z^{4}+2z^{3}+}{\color{blue}0z^{2}+0z^{1}+10+0z^{-1}+0z^{-2}}{\color{red}+2z^{-3}-2z^{-4}}     \\
{\color{red}0z^{4}+0z^{3}+}{\color{blue}0z^{2}+0z^{1}+0+0z^{-1}+0z^{-2}}{\color{red}+0z^{-3}+0z^{-4}}
\end{matrix}\right. \nonumber \\
&\qquad \qquad \qquad \qquad \qquad \qquad \left.\begin{matrix}
 {\color{red}0z^{4}+0z^{3}+}{\color{blue}0z^{2}+0z^{1}+0+0z^{-1}+0z^{-2}}{\color{red}+0z^{-3}+0z^{-4}} \\
{\color{red}-2z^{4}+2z^{3}+}{\color{blue}0z^{2}+0z^{1}+10+0z^{-1}+0z^{-2}}{\color{red}+2z^{-3}-2z^{-4}}
\end{matrix}\right]_{2\times 2}.
\end{align} By applying the zone extraction function for the time-shifts from $-2$ to $2$, i.e., $Z=3$, we have
\begin{align}
f_{Z}\left(\widetilde{\textbf{X}(z)}\cdot\textbf{X}(z)\right)&=\Big[f_{Z}\left(S_{\textbf{x}_\mu,\textbf{x}_\nu}(z)\right)\Big]_{2\times 2} \nonumber \\
&=\begin{bmatrix}
{\color{blue}0z^{2}+0z^{1}+10+0z^{-1}+0z^{-2}}& {\color{blue}0z^{2}+0z^{1}+0+0z^{-1}+0z^{-2}}\\
{\color{blue}0z^{2}+0z^{1}+0+0z^{-1}+0z^{-2}} & {\color{blue}0z^{2}+0z^{1}+10+0z^{-1}+0z^{-2}}
\end{bmatrix}_{2\times 2} \nonumber \\
&=\begin{bmatrix}
{\color{blue}10}& {\color{blue}0}\\
{\color{blue}0} & {\color{blue}10}
\end{bmatrix}_{2\times 2}= 10 \cdot\textbf{I}_2
\end{align} for the ZCZ width $Z=3$  and hence $\textbf{X}(z)$ is a binary $3$-PU matrix of size $2\times 2$ and sequence length $5$.
\end{example}
\begin{example}
\label{Example:5-PU_4_2}
Let $M=2, K=4$ and $L=16$. A binary polynomial matrix $\textbf{X}(z)$ of size $2\times 4$ and length $L=16$ is given by (\ref{4_2_16_5_PU}) in which $+$ and $-$ denote $+1$ and $-1$, respectively. Then, the matrix of ACCF sums is given by
\begin{align}
&\widetilde{\textbf{X}(z)}\cdot\textbf{X}(z)=\Big[S_{\textbf{x}_\mu, \textbf{x}_\nu}(z)
\Big]_{4\times 4} =  \left[{\color{blue}32\cdot \delta(\mu-\nu)}{\color{red}+\sum_{5\leq |\tau|<16}S_{\textbf{x}_\mu, \textbf{x}_\nu}[\tau]\cdot z^{-\tau}}\right]_{4\times 4}.
\end{align} Thus, $f_{Z}\left(\widetilde{\textbf{X}(z)}\cdot\textbf{X}(z)\right)=32\cdot \textbf{I}_4$ for $Z=5$  and hence $\textbf{X}(z)$ is a binary $5$-PU matrix of size $2\times 4$ and sequence length $16$. Note that a  $2\times 4$ PU matrix does not exist.
	\begin{equation}
	\label{4_2_16_5_PU}
	\textbf{X}^T=\begin{bmatrix}
	++-+++-+---+---+ & --+-++-++++----+  \\
	--+-++-++++----+ & ++-+++-+---+---+ \\
	+---+----+---+-- & -++++---+-++-+-- \\
	-++++---+-++-+-- & +---+----+---+--
	\end{bmatrix}.
	\end{equation}
\end{example}
\section{\textbf{A Unifying Construction Framework for ZPU Matrices}}
\label{Unifying:Framework:sec:pro}
In this section, we will first investigate a unifying construction framework for ZPU matrices. We show that ZPU matrices include existing PU matrices as a special case. Then, we show that optimal ZPU matrices can also be generated from our proposed unifying construction framework.
\subsection{\textbf{Proposed Unifying Construction Framework for ZPU Matrices}}
\label{subsection:unifying:construction} In this subsection, we propose a construction method for ZPU matrices based on block matrices. Then, we discuss on lengths and phases of the constructed sequences.
\subsubsection{\textbf{Proposed Construction}}
\begin{theorem}
\label{theorem:seed:Z:PU:unifying}
Let $M$ and $K$ be two positive integers such that $K=MP$ for some positive integer $P$. Let $\bm{\mathcal{G}}^{(0)}(z)$ and $\bm{\mathcal{G}}^{(1)}(z)$ be two PU matrices of size $K\times K$, $M\times M$ and sequence length $L_0$, $L_1$, respectively. Then, a polynomial matrix $\bm{\mathcal{G}}(z)$ of size $M\times K$ and sequence length $KL_0L_1$ is given by
\begin{equation}
\label{Z-PU:unifying}
\bm{\mathcal{G}}(z)=\left(\textbf{1}_{P}\otimes \bm{\mathcal{G}}^{(1)}\left(z^{KL_0}\right)\right)\cdot \textbf{D}_K\left(z^{\alpha L_0}\right)\cdot \bm{\mathcal{G}}^{(0)}\left(z^{\beta K}\right),
\end{equation} where $\textbf{D}_K(z)=\text{diag}\Big(1,z^{-1}, \cdots, z^{-(K-1)}\Big)$, $\textbf{1}_{P}$ is an $1\times P$ row vector with all ones, $\otimes$ denotes the Kronecker product and $(\alpha , \beta)=\left(1/L_0,1\right)$ or $\left(1, 1/K\right)$. Then, the matrix $\bm{\mathcal{G}}(z)$ is a unimodular ZPU matrix of size $M\times K$, sequence length $KL_0L_1$ and ZCZ width $Z=M$.
\end{theorem}
\begin{IEEEproof} To prove \textit{Theorem \ref{theorem:seed:Z:PU:unifying}}, we carry out the proof into the following two cases.
\newline
\textbf{Case-I $\left[(\alpha , \beta)=\left(1/L_0,1\right)\right]$}: Let $\bm{\mathcal{G}}^{(0)}\left(z\right)=\left[g^{(0)}_{\mu\nu}(z)\right]_{K\times K},\bm{\mathcal{G}}^{(1)}\left(z\right)=\left[g^{(1)}_{mn}(z)\right]_{M\times M}$, and $\textbf{1}_{P}\otimes \bm{\mathcal{G}}^{(1)}\left(z\right)=\left[h_{mk}(z)\right]_{M\times K}$. Let $\textbf{g}^{(0)}_{\nu}(z)$, $\textbf{g}^{(1)}_{n}(z)$, and $\textbf{h}_{k}(z)$  be the $\nu$-th, $n$-th, and $k$-th column vectors of the matrix $\bm{\mathcal{G}}^{(0)}\left(z\right), \bm{\mathcal{G}}^{(1)}\left(z\right)$ and $\textbf{1}_{P}\otimes \bm{\mathcal{G}}^{(1)}\left(z\right)$, respectively. Note that $\textbf{1}_{P}\otimes \bm{\mathcal{G}}^{(1)}\left(z\right)$ is an $M\times K$ block polynomial matrix with $P$ identical matrix blocks. Thus, we have $\textbf{h}_{pM+m}(z)=\textbf{g}^{(1)}_{m}(z)$ for $p=0,1,\cdots, P-1$ and $m=0,1,\cdots, M-1$. Since $\bm{\mathcal{G}}^{(0)}\left(z\right)$ and $\bm{\mathcal{G}}^{(1)}\left(z\right)$ are PU matrices, we can write
\begin{align*}
\widetilde{\textbf{g}^{(0)}_{\nu}(z)}\cdot \textbf{g}^{(0)}_{\mu}(z) =KL_0\cdot \delta(\mu-\nu), \quad \text{and}\quad  \widetilde{\textbf{g}^{(1)}_{m'}(z)}\cdot \textbf{g}^{(1)}_{m}(z) =ML_1\cdot \delta(m'-m).
\end{align*} Consequently, we have
\begin{align}
\label{unifying:Matrix}
\widetilde{\textbf{h}_{p'M+m'}(z)}\cdot \textbf{h}_{pM+m}(z)&=\widetilde{\textbf{g}^{(1)}_{m'}(z)}\cdot \textbf{g}^{(1)}_{m}(z) =ML_1\cdot \delta(m'-m).
\end{align} According to (\ref{Z-PU:unifying}), the $\mu$-th column of the matrix $\bm{\mathcal{G}}(z)$ is given by
\begin{equation}
\textbf{g}_\mu(z)=\Big[g_{0\mu}(z),  g_{1\mu}(z), \cdots,  g_{(M-1)\mu}(z)\Big]^T,
\end{equation} where $\mu=0,1,\cdots, K-1$ and the sequence $g_{n\mu}(z)$ of length $KL_0L_1$ can be written as
\begin{align}
\label{unifying:entry_G}
g_{n\mu}(z)&=\sum_{m=0}^{M-1}\sum_{p=0}^{P-1}h_{n(pM+m)}\left(z^{KL_0}\right)\cdot g^{(0)}_{(pM+m)\mu}\left(z^{K}\right)\cdot z^{-(pM+m)} \nonumber \\
&=\sum_{m=0}^{M-1}\sum_{p=0}^{P-1}\sum_{l_0=0}^{L_0-1}\sum_{l_1=0}^{L_1-1}h_{n(pM+m)}\left[l_1\right]\cdot g^{(0)}_{(pM+m)\mu}\left[l_0\right]\cdot z^{-\left(pM+m+(l_0+l_1L_0)K\right)},
\end{align} where $n=0,1,\cdots, M-1$. Equivalently, we can write (\ref{unifying:entry_G}) in time-domain as follows:
\begin{equation}
\label{time:domain:expres:unify}
g_{n\mu}[k]=h_{n(pM+m)}\left[l_1\right]\cdot g^{(0)}_{(pM+m)\mu}\left[l_0\right],
\end{equation} where $k=pM+m+(l_0+l_1L_0)K$. Since the product of unimodular complex numbers is unimodular, each sequence $g_{n\mu}[k]$ is a unimodular sequence of length $KL_0L_1$. We now consider the sum of ACCFs. The sum of ACCFs between the $\mu$-th and $\nu$-th columns of the matrix $\bm{\mathcal{G}}(z)$ is given by
\begin{align}
&S_{\textbf{g}_\mu,\textbf{g}_\nu}(z)=\sum_{n=0}^{M-1}R_{\textbf{\textit{g}}_{n\mu}, \textbf{\textit{g}}_{n\nu}}(z) \nonumber \\
&=\sum_{n=0}^{M-1}g_{n\mu}(z^{-1})\cdot g_{n\nu}^{*}(z) \nonumber \\
&=\sum_{n=0}^{M-1}\sum_{m=0}^{M-1}\sum_{m'=0}^{M-1}\sum_{p=0}^{P-1}\sum_{p'=0}^{P-1} h_{n(pM+m)}\left(z^{-KL_0}\right)\cdot g^{(0)}_{(pM+m)\mu}\left(z^{-K}\right)\cdot h_{n(p'M+m')}^*\left(z^{KL_0}\right)\cdot g^{(0)*}_{(p'M+m')\nu}\left(z^{K}\right) \nonumber \\
&\qquad \qquad \qquad \qquad \qquad  \cdot z^{-\{(p'-p)M+(m'-m)\}} \quad (\text{Using (\ref{unifying:entry_G})}) \nonumber \\
&=\sum_{m=0}^{M-1}\sum_{m'=0}^{M-1}\sum_{p=0}^{P-1}\sum_{p'=0}^{P-1} \left\{\sum_{n=0}^{M-1}h_{n(pM+m)}\left(z^{-KL_0}\right)\cdot h_{n(p'M+m')}^*\left(z^{KL_0}\right)\right\}\cdot  g^{(0)}_{(pM+m)\mu}\left(z^{-K}\right)\cdot  g^{(0)*}_{(p'M+m')\nu}\left(z^{K}\right) \nonumber \\
&\qquad \qquad \qquad \qquad \qquad \cdot z^{-\{(p'-p)M+(m'-m)\}}  \nonumber \\
&=\sum_{m=0}^{M-1}\sum_{m'=0}^{M-1}\sum_{p=0}^{P-1}\sum_{p'=0}^{P-1} \left\{\widetilde{\textbf{h}_{p'M+m'}\left(z^{-KL_0}\right)}\cdot \textbf{h}_{pM+m}\left(z^{-KL_0}\right)\right\}\cdot  g^{(0)}_{(pM+m)\mu}\left(z^{-K}\right)\cdot  g^{(0)*}_{(p'M+m')\nu}\left(z^{K}\right) \nonumber \\
&\qquad \qquad \qquad \qquad \qquad \cdot z^{-\{(p'-p)M+(m'-m)\}}  \nonumber \\
&=\sum_{m=0}^{M-1}\sum_{m'=0}^{M-1}\sum_{p=0}^{P-1}\sum_{p'=0}^{P-1} \Big\{ML_1\cdot \delta(m'-m)\Big\}\cdot  g^{(0)}_{(pM+m)\mu}\left(z^{-K}\right)\cdot  g^{(0)*}_{(p'M+m')\nu}\left(z^{K}\right)  \quad (\text{Using (\ref{unifying:Matrix})}) \qquad  \nonumber \\
&\qquad \qquad \qquad \qquad \qquad \cdot  z^{-\{(p'-p)M+(m'-m)\}} \nonumber \\
&=ML_1\sum_{m=0}^{M-1}\sum_{p=0}^{P-1}\sum_{p'=0}^{P-1}  g^{(0)}_{(pM+m)\mu}\left(z^{-K}\right)\cdot  g^{(0)*}_{(p'M+m)\nu}\left(z^{K}\right) \cdot z^{-(p'-p)M} \nonumber \\
&={\color{blue}ML_1\sum_{m=0}^{M-1}\sum_{p=0}^{P-1}g^{(0)}_{(pM+m)\mu}\left(z^{-K}\right)\cdot  g^{(0)*}_{(pM+m)\nu}\left(z^{K}\right)} \nonumber \\
& \quad {\color{red}+ML_1\sum_{m=0}^{M-1}\sum_{p=0}^{P-1}\sum_{p'=0, p'\neq p}^{P-1}  g^{(0)}_{(pM+m)\mu}\left(z^{-K}\right)\cdot  g^{(0)*}_{(p'M+m)\nu}\left(z^{K}\right) \cdot z^{-(p'-p)M}} \nonumber \\
&={\color{blue}ML_1\cdot \widetilde{\textbf{g}^{(0)}_{\nu}\left(z^{-K}\right)}\cdot  \textbf{g}^{(0)}_{\mu}\left(z^{-K}\right)}   \nonumber \\
&\quad {\color{red}+ML_1\sum_{m=0}^{M-1}\sum_{p=0}^{P-1}\sum_{p'=0, p'\neq p}^{P-1}  g^{(0)}_{(pM+m)\mu}\left(z^{-K}\right)\cdot  g^{(0)*}_{(p'M+m)\nu}\left(z^{K}\right) \cdot z^{-(p'-p)M}} \nonumber \\
&={\color{blue}MKL_0L_1 \cdot \delta(\mu-\nu)}{\color{red}+ML_1\sum_{m=0}^{M-1}\sum_{p=0}^{P-1}\sum_{p'=0, p'\neq p}^{P-1}  g^{(0)}_{(pM+m)\mu}\left(z^{-K}\right)\cdot  g^{(0)*}_{(p'M+m)\nu}\left(z^{K}\right) \cdot z^{-(p'-p)M}}
\end{align} The matrix of ACCF sums between sets is given by
\begin{align}
&\widetilde{\bm{\mathcal{G}}(z)}\cdot\bm{\mathcal{G}}(z)=\begin{bmatrix}
S_{\textbf{g}_\mu,\textbf{g}_\nu}(z)
\end{bmatrix}_{K\times K} =\left[{\color{blue}MKL_0L_1\cdot \delta(\mu-\nu)}{\color{red}+\sum_{M\leq |\tau|<KL_0L_1}S_{\textbf{g}_\mu,\textbf{g}_\nu}[\tau]\cdot z^{-\tau}}\right]_{K\times K}.
\end{align} Thus, we have
\begin{align}
f_{Z}\left(\widetilde{\bm{\mathcal{G}}(z)}\cdot \bm{\mathcal{G}}(z)\right)&=\Big[f_{Z}\left(S_{\textbf{g}_\mu,\textbf{g}_\nu}(z)\right)\Big]_{K\times K} \nonumber \\
&= \Big[{\color{blue}MKL_0L_1\cdot \delta(\mu-\nu)}\Big]_{K\times K} \nonumber \\
&=MKL_0L_1\cdot \textbf{I}_K\ \text{for the ZCZ width $Z=M$}.
\end{align} That is, $\bm{\mathcal{G}}(z)$  is a ZPU matrix of size $M\times K$, sequence length $KL_0L_1$ and ZCZ width $Z=M$.

\textbf{Case-II $\left[(\alpha , \beta)=\left(1, 1/K\right)\right]$}: This case can be proved with the similar approach to the above case. This completes the proof.
\end{IEEEproof}
\begin{remark}
Note that the proposed construction framework corresponds to the interleaving and concatenation of the sequences from $\bm{\mathcal{G}}^{(0)}\left(z\right)$ if we consider the cases when $(\alpha , \beta)=\left(1/L_0,1\right)$ and $(\alpha , \beta)=\left(1, 1/K\right)$, respectively.
\end{remark}
  According to (\ref{Z-PU:unifying}), we have the following corollary.
\begin{corollary}
\label{cor:PU:Z_PU:unifying}
The polynomial matrix $\bm{\mathcal{G}}(z)$ given by (\ref{Z-PU:unifying}) becomes a PU matrix with size $K\times K$ and sequence length $KL_0L_1$ when $P=1$. More specifically, when $P=1$, we can write
\begin{equation}
\bm{\mathcal{G}}(z)= \bm{\mathcal{G}}^{(1)}\left(z^{KL_0}\right)\cdot \textbf{D}_K\left(z^{\alpha L_0}\right)\cdot \bm{\mathcal{G}}^{(0)}\left(z^{\beta K}\right).
\end{equation} In this case, $\bm{\mathcal{G}}(z)$ is a $K\times K$ PU matrix with sequence length $KL_0L_1$.
\end{corollary}

\begin{remark}
Based on \textit{Corollary \ref{cor:PU:Z_PU:unifying}}, we can say that the proposed unifying framework includes PU generating matrices as a special case when $P=1$.
\end{remark}
\begin{remark}
Our previous construction framework \cite[eq.(19)]{2019ShibsankarTSP} suggests that it is possible to construct ZPU matrix with various sequence lengths. 
\end{remark}
We illustrate our proposed construction of polyphase ZPU matrix by the following example.
\begin{example}
Let $M=3, P=2$ and $K=6$. Let $\bm{\mathcal{G}}^{(0)}(z)$ be a PU matrix with size $6\times 6$ and sequence length $L_0=2$ given in \textit{Example \ref{example:SPL:2018:Flexible}}. Let $\bm{\mathcal{G}}^{(1)}\left(z\right)=\textbf{F}_3\cdot \textbf{D}_3(z) \cdot \textbf{F}_3$ be a PU matrix with size $3\times 3$ and sequence length $L_1=3$, where $\textbf{D}_3(z)=\text{diag}\left(1,z^{-1}, z^{-2}\right)$. Let $(\alpha, \beta)=\left(1/L_0,1\right)=(1/2,1)$. We have $KL_0L_1=36$, $Z=M=3$ and $MKL_0L_1=108$. According to (\ref{Z-PU:unifying}), a $3$-PU matrix with size $3\times 6$ and sequence length $36$ is given by
\begin{align}
\bm{\mathcal{G}}(z)&= \left(\textbf{1}_{P}\otimes \bm{\mathcal{G}}^{(1)}\left(z^{KL_0}\right)\right)\cdot \textbf{D}_K\left(z^{\alpha L_0}\right)\cdot \bm{\mathcal{G}}^{(0)}\left(z^{\beta K}\right) \nonumber \\
&=\left[\bm{\mathcal{G}}^{(1)}\left(z^{12}\right) \ \bm{\mathcal{G}}^{(1)}\left(z^{12}\right)\right]\cdot \textbf{D}_6\left(z\right)\cdot \bm{\mathcal{G}}^{(0)}(z^{6}),
\end{align} where $\textbf{D}_6\left(z\right)=\text{diag}\Big(1, z^{-1}, z^{-2},z^{-3}, z^{-4}, z^{-5}\Big)$. The matrix of ACCF sums is given by
\begin{align}
&\widetilde{\bm{\mathcal{G}}(z)}\cdot\bm{\mathcal{G}}(z)=\Big[S_{\textbf{g}_\mu,\textbf{g}_\nu}(z)
\Big]_{6\times 6} =\left[{\color{blue}108\cdot \delta(\mu-\nu)}{\color{red}+\sum_{3\leq |\tau|<36}S_{\textbf{g}_\mu,\textbf{g}_\nu}[\tau]\cdot z^{-\tau}}\right]_{6\times 6}.
\end{align} Thus, $f_{Z}\left(\widetilde{\bm{\mathcal{G}}(z)}\cdot\bm{\mathcal{G}}(z)\right)=108\cdot \textbf{I}_6$ for the ZCZ width $Z=3$.  We have written out this $3$-PU matrix by Table \ref{Table:3-PU:Matrix:L36} in which only the exponents of $\omega=e^{-2\pi \sqrt{-1}/6}$ are given. The number of phases of the constructed sequence is $LCM\{3,6\}=6$ as the number of phases of $\bm{\mathcal{G}}^{(0)}(z)$ and $\bm{\mathcal{G}}^{(1)}(z)$ are $6$ and $3$, respectively. The $(n, \mu)$-th entry (i.e., $(n,\mu)$-th sequence) of $\bm{\mathcal{G}}(z)$ can be written in time-domain as follows
 \begin{equation}
\label{example:time:domain:expres:unify}
g_{n\mu}[k]=h_{n(pM+m)}\left[l_1\right]\cdot g^{(0)}_{(pM+m)\mu}\left[l_0\right],
\end{equation} where $k=3p+m+(l_0+2l_1)6$ with $p,l_0=0,1$ and $m,l_1=0,1,2$.
\end{example}

\subsubsection{\textbf{Lengths and Phases of the Constructed Sequences}}
Besides the zero auto- and cross-correlation properties within the ZCZ width of ZCCS, sequence
lengths and phases play an important role in many practical scenarios. We now discuss on sequence lengths and phases of the constructed matrices.

\textit{\textbf{Sequence Lengths:}} The constructed matrix $\bm{\mathcal{G}}(z)$ from \textit{Theorem \ref{theorem:seed:Z:PU:unifying}} consists of $K$ distinct sequence sets (columns) and each set consists of $M$ unimodular sequences with identical length $KL_0L_1$. By applying \textit{Lemma \ref{SPL:2018}}, we have sequence lengths $L_0=d^{N_0}_0$ for the matrix $\bm{\mathcal{G}}^{(0)}(z)$ and $L_1=d^{N_1}_1$ for the matrix $\bm{\mathcal{G}}^{(1)}(z)$, where $d_0|K$, $d_1|M$ and $N_0,N_1 \in \mathbb{N}$. Therefore, we can construct unimodular ZPU matrices with sequence lengths $Kd^{N_0}_0d^{N_1}_1$.

\textit{\textbf{Phases:}} According to our proposed construction framework (\ref{Z-PU:unifying}), the choices of unitary matrices play an important role in determining phases, set sizes, ZCZ widths and sequence lengths of the constructed ZPU matrix $\bm{\mathcal{G}}(z)$. From Table \ref{Table:BH:Matrices}, there are many distinct $BH$ matrices with distinct phases for the given matrix size. For example, there are $BH(4,2)$, $BH(4,4)$, and $BH(4,6)$ with matrix size $4\times 4$. Therefore, the availability of wider range of unitary matrices enables (\ref{Z-PU:unifying}) to produce many new ZCCSs compared to the existing techniques. The number of phases for the constructed sequence is $q=LCM\{q_0,q_1\}$, where the number of phases of $\bm{\mathcal{G}}^{(0)}(z)$ and $\bm{\mathcal{G}}^{(1)}(z)$ are $q_0$-th and $q_1$-th root of unity, respectively, with $2\leq q_0\leq K$ and $2\leq q_1\leq M$. Therefore, phases of the constructed sequences can be controlled by the appropriate choice of $\bm{\mathcal{G}}^{(0)}(z)$ and $\bm{\mathcal{G}}^{(1)}(z)$ for the given positive integers $K$ and $M$. For example, binary sequences (i.e., $q=2$) can be constructed when both the matrices $\bm{\mathcal{G}}^{(0)}(z)$ and $\bm{\mathcal{G}}^{(1)}(z)$ have phases $q_0=q_1=2$.

\begin{center}
\captionof{table}{A $3$-PU Matrix with Size $3 \times 6$ and Sequence Length $36$}  \label{Table:3-PU:Matrix:L36}
	%\resizebox{\linewidth}{!}{
		\renewcommand{\arraystretch}{1.3}
\begin{tabular}{|l|l|l|l|}
\hline
\multicolumn{1}{|c|}{} & \multicolumn{1}{c||}{$  0     0     0     0     0     0     0     3     0     3     0     3     0     2     4     0     2     4     0     5     4     3     2     1     0     4     2     0     4     2     0     1     2     3     4     5$} & \multicolumn{1}{c|}{} & \multicolumn{1}{c|}{$     0     0     2     2     4     4     0     3     2     5     4     1     0     2     0     2     0     2     0     5     0     5     0     5     0     4     4     2     2     0     0     1     4     5     2     3$} \\
\multicolumn{1}{|c|}{$\textbf{g}_0$} & \multicolumn{1}{c||}{$    0     0     0     0     0     0     0     3     0     3     0     3     2     4     0     2     4     0     2     1     0     5     4     3     4     2     0     4     2     0     4     5     0     1     2     3$} & \multicolumn{1}{c|}{$\textbf{g}_1$} & \multicolumn{1}{c|}{$    0     0     2     2     4     4     0     3     2     5     4     1     2     4     2     4     2     4     2     1     2     1     2     1     4     2     2     0     0     4     4     5     2     3     0     1$} \\
\multicolumn{1}{|c|}{} & \multicolumn{1}{c||}{$   0     0     0     0     0     0     0     3     0     3     0     3     4     0     2     4     0     2     4     3     2     1     0     5     2     0     4     2     0     4     2     3     4     5     0     1$} & \multicolumn{1}{c|}{} & \multicolumn{1}{c|}{$    0     0     2     2     4     4     0     3     2     5     4     1     4     0     4     0     4     0     4     3     4     3     4     3     2     0     0     4     4     2     2     3     0     1     4     5$} \\
\hline
\hline
\multicolumn{1}{|c|}{} & \multicolumn{1}{c||}{$0     0     0     0     4     4     2     5     4     1     2     5     0     2     4     0     0     2     2     1     2     1     4     3     0     4     2     0     2     0     2     3     0     1     0     1$} & \multicolumn{1}{c|}{} & \multicolumn{1}{c|}{$  0     0     4     4     2     2     2     5     0     3     4     1     0     2     2     4     4     0     2     1     4     3     0     5     0     4     0     4     0     4     2     3     2     3     2     3$} \\
\multicolumn{1}{|c|}{$\textbf{g}_2$} & \multicolumn{1}{c||}{$    0     0     0     0     4     4     2     5     4     1     2     5     2     4     0     2     2     4     4     3     4     3     0     5     4     2     0     4     0     4     0     1     4     5     4     5$} & \multicolumn{1}{c|}{$\textbf{g}_3$} & \multicolumn{1}{c|}{$   0     0     4     4     2     2     2     5     0     3     4     1     2     4     4     0     0     2     4     3     0     5     2     1     4     2     4     2     4     2     0     1     0     1     0     1$} \\
\multicolumn{1}{|c|}{} & \multicolumn{1}{c||}{$    0     0     0     0     4     4     2     5     4     1     2     5     4     0     2     4     4     0     0     5     0     5     2     1     2     0     4     2     4     2     4     5     2     3     2     3$} & \multicolumn{1}{c|}{} & \multicolumn{1}{c|}{$ 0     0     4     4     2     2     2     5     0     3     4     1     4     0     0     2     2     4     0     5     2     1     4     3     2     0     2     0     2     0     4     5     4     5     4     5$} \\
\hline
\hline
\multicolumn{1}{|c|}{} & \multicolumn{1}{c||}{$     0     0     4     4     0     0     4     1     2     5     2     5     0     2     2     4     2     4     4     3     0     5     4     3     0     4     0     4     4     2     4     5     4     5     0     1$} & \multicolumn{1}{c|}{} & \multicolumn{1}{c|}{$0     0     2     2     2     2     4     1     4     1     0     3     0     2     0     2     4     0     4     3     2     1     2     1     0     4     4     2     0     4     4     5     0     1     4     5$} \\
\multicolumn{1}{|c|}{$\textbf{g}_4$} & \multicolumn{1}{c||}{$     0     0     4     4     0     0     4     1     2     5     2     5     2     4     4     0     4     0     0     5     2     1     0     5     4     2     4     2     2     0     2     3     2     3     4     5$} & \multicolumn{1}{c|}{$\textbf{g}_5$} & \multicolumn{1}{c|}{$    0     0     2     2     2     2     4     1     4     1     0     3     2     4     2     4     0     2     0     5     4     3     4     3     4     2     2     0     4     2     2     3     4     5     2     3$} \\
\multicolumn{1}{|c|}{} & \multicolumn{1}{c||}{$  0     0     4     4     0     0     4     1     2     5     2     5     4     0     0     2     0     2     2     1     4     3     2     1     2     0     2     0     0     4     0     1     0     1     2     3$} & \multicolumn{1}{c|}{} & \multicolumn{1}{c|}{$    0     0     2     2     2     2     4     1     4     1     0     3     4     0     4     0     2     4     2     1     0     5     0     5     2     0     0     4     2     0     0     1     2     3     0     1$} \\
\hline
\end{tabular}
\end{center}

\subsection{\textbf{Optimal Seed ZPU Matrices}}
\label{sub:sec:optimal:ZPU}
In this subsection, we propose a novel construction of optimal seed ZPU matrices. Then, we propose a new construction of optimal ZPU matrices with  larger ZCZ widths by using these seed ZPU matrices in the subsequent subsection.

Let $M$ and $K$ be two positive integers such that $K=MP$ for some positive integer $P$. Let $\textbf{U}_K$ and $\textbf{U}_{M}$ be two $BH$ matrices of size $K\times K$ and $M\times M$, respectively. Let us consider a block matrix $\textbf{G}=\textbf{1}_{P}\otimes \textbf{U}_M$. Clearly, $\textbf{G}$ is a matrix of size $M\times K$. Then, a polynomial matrix $\bm{\mathcal{G}}(z)$ of size $M\times K$ and degree $K-1$ is given by
\begin{equation}
\label{Seed:Z-PU:non:power:two}
\bm{\mathcal{G}}(z)=\left\{\textbf{1}_{P}\otimes \textbf{U}_M\right\}\cdot \textbf{D}_K(z)\cdot \textbf{U}_K.
\end{equation}
\begin{corollary}
\label{corollary:seed:Z:PU}
The matrix $\bm{\mathcal{G}}(z)$ given by (\ref{Seed:Z-PU:non:power:two}) is an optimal unimodular ZPU matrix of size $M\times K$, sequence length $K$ and ZCZ width $Z=M$.
\end{corollary}
\begin{IEEEproof}
According to \textit{Theorem \ref{theorem:seed:Z:PU:unifying}}, $\bm{\mathcal{G}}(z)$  is a ZPU matrix of size $M\times K$ and sequence length $K$ and ZCZ width $Z=M$ by using $BH$ matrices instead of PU matrices $\bm{\mathcal{G}}^{(0)}(z)$ and $\bm{\mathcal{G}}^{(1)}(z)$. Note that any arbitrary $BH$ matrix can be considered as a square PU matrix of sequence length $1$. In addition, we have $K/M=MP/M=P=\lfloor{L/Z}\rfloor$. So, $\bm{\mathcal{G}}(z)$ is an optimal $M$-PU matrix. This completes the proof.
\end{IEEEproof} We illustrate our proposed construction of optimal seed ZPU matrices by the following example. We give a new $3$-PU matrix of size $3\times 6$ and sequence length $6$ with $3$-phase-shift keying (PSK) constellation as opposed to the case when we will use DFT matrix where the generated sequences belong to the $6$-PSK constellation.

\begin{example}
\label{example:seed:Z-PU:Matrix:M3}
Let $M=3$ and $K=6$ for $P=2$. Let $\textbf{U}_K=\textbf{S}_{6}=BH(6,3)$ given by (\ref{BH(6,3):spectral:matrix}) in \textit{Example \ref{BH(6,3):example}}. Let $\textbf{U}_M=\textbf{F}_3$ and $\textbf{G}=\begin{bmatrix}
\textbf{F}_3 & \textbf{F}_3
\end{bmatrix}$. Clearly, $\textbf{G}$ is a $3\times 6$ matrix with $3$-PSK constellation. According to our proposed construction method, a $3$-PU matrix of size $3\times 6$ and sequence length $6$ is given by
\begin{align}
\label{Example:Seed:Z-PU:non:power:two}
\bm{\mathcal{G}}(z)&=\textbf{G}\cdot \textbf{D}_6(z)\cdot \textbf{U}_K= \begin{bmatrix}
\textbf{F}_3 & \textbf{F}_3
\end{bmatrix}\cdot \textbf{D}_6(z)\cdot \textbf{S}_6.
\end{align} The matrix of ACCF sums is given by
\begin{align}
&\widetilde{\bm{\mathcal{G}}(z)}\cdot\bm{\mathcal{G}}(z)=\Big[S_{\textbf{g}_\mu,\textbf{g}_\nu}(z)
\Big]_{6\times 6} =\left[{\color{blue}18\cdot \delta(\mu-\nu)}{\color{red}+\sum_{3\leq |\tau|<6}S_{\textbf{g}_\mu,\textbf{g}_\nu}[\tau]\cdot z^{-\tau}}\right]_{6\times 6}.
\end{align} Thus, $f_{Z}\left(\widetilde{\bm{\mathcal{G}}(z)}\cdot\bm{\mathcal{G}}(z)\right)=18\cdot \textbf{I}_6$ for the ZCZ width $Z=3$. Also, we have $K/M=P=\lfloor{L/Z}\rfloor$ and hence $\bm{\mathcal{G}}(z)$ is an optimal $3$-PU matrix of size $3\times 6$ and sequence length $6$ with $3$-PSK constellation. We have written out this $3$-PU matrix by Table \ref{Table:Z-PU:3-PU:Matrix:L6} in which only the exponents of $\omega=e^{-2\pi \sqrt{-1}/3}$ are given.
\end{example}
\begin{center}
\captionof{table}{A New $3 \times 6$ Optimal $3$-PU Matrix with Sequence Length $6$}  \label{Table:Z-PU:3-PU:Matrix:L6}
	%\resizebox{\linewidth}{!}{
		\renewcommand{\arraystretch}{1.05}
\begin{tabular}{|l|l||l|l||l|l||l|l||l|l||l|l|}
\hline
 & $000000$ &  & $0     0     1     1     2     2$ &  & $0     1     0     2     2     1$ &  & $0     1     2     0     1     2$ &  & $ 0     2     2     1     0     1$ &  & $0     2     1     2     1     0$ \\
$\textbf{x}_0$ & $012012$ & $\textbf{x}_1$ & $0     1     0     1     0     1$ & $\textbf{x}_2$ & $0     2     2     2     0     0$ & $\textbf{x}_3$ & $0     2     1     0     2     1$ & $\textbf{x}_4$ & $ 0     0     1     1     1     0$ & $\textbf{x}_5$ & $0     0     0     2     2     2$ \\
 & $021021$ &  & $ 0     2     2     1     1     0$ &  & $0     0     1     2     1     2$ &  & $0     0     0     0     0     0$ &  & $0     1     0     1     2     2$ &  & $0     1     2     2     0     1$ \\
\hline
\end{tabular}
\end{center}

According to \textit{Corollary \ref{corollary:seed:Z:PU}}, we have the following result for the construction of binary sequences.
\begin{corollary}
\label{cor:theorem2}
Let $\textbf{H}_{2^{m+k}}$ and $\textbf{H}_{2^{m}}$ be two binary Hadamard matrices of size $2^{m+k}\times 2^{m+k}$ and $2^{m}\times 2^{m}$, respectively, for some positive integers $m$ and $k$. Let us consider a block matrix $\textbf{G}=\left[\textbf{H}_{2^{m}}\right.$ $ \ \textbf{H}_{2^{m}}$ $\ \cdots$ $\  \textbf{H}_{2^{m}}$ $\left.(\text{$2^k$ times})\right]$ with size $2^{m}\times 2^{m+k}$. Then, the matrix $\bm{\mathcal{G}}(z)$ given by (\ref{Seed:Z-PU:non:power:two}) is an optimal binary $2^m$-PU matrix of size $2^{m}\times 2^{m+k}$ and sequence length $2^{m+k}$. That is, $\bm{\mathcal{G}}(z)$ represents an optimal binary $(2^{m+k},2^{m})$-ZCCS$^{2^{m+k}}_{2^{m}}$.
\end{corollary} We give the following example to illustrate the above corollary.
\begin{example}
\label{example:seed:Z-PU:Matrix}
Let $M=2$ and $K=4$ for $P=2$. Let $\textbf{U}_K=\textbf{H}_4$ and $\textbf{U}_M=\textbf{H}_2$. Let $\textbf{G}=\begin{bmatrix}
\textbf{H}_2 & \textbf{H}_2
\end{bmatrix}$. Clearly, $\textbf{G}$ is a $2\times 4$ binary matrix. According to our proposed construction method, a binary $2$-PU matrix of size $2\times 4$ and sequence length $4$ is given by
\begin{align}
\label{Example:Seed:Z-PU:2^m}
\bm{\mathcal{G}}(z)&=\Big[\textbf{H}_2 \ \ \textbf{H}_2
\Big]\cdot \textbf{D}_4(z)\cdot \textbf{H}_4 \nonumber \\
&\equiv \begin{bmatrix}
++++ &+-+-&++--&+--+\\
+-+-&++++&+--+&++--
\end{bmatrix},
\end{align} where $\textbf{D}_4(z)=\text{diag}\Big(1,z^{-1}, z^{-2}, z^{-3}\Big)$. The matrix of ACCF sums is given by
\begin{align}
\widetilde{\bm{\mathcal{G}}(z)}\cdot\bm{\mathcal{G}}(z)&=\Big[S_{\textbf{g}_\mu,\textbf{g}_\nu}(z)
\Big]_{4\times 4}  =\left[{\color{blue}8\cdot \delta(\mu-\nu)}{\color{red}+\sum_{2\leq |\tau|<4}S_{\textbf{g}_\mu,\textbf{g}_\nu}[\tau]\cdot z^{-\tau}}\right]_{4\times 4}.
\end{align} Thus, $f_{Z}\left(\widetilde{\bm{\mathcal{G}}(z)}\cdot\bm{\mathcal{G}}(z)\right)=8\cdot \textbf{I}_4$ for the ZCZ width $Z=2$. Also, we have $K/M=P=\lfloor{L/Z}\rfloor$ and hence $\bm{\mathcal{G}}(z)$ is an optimal binary $2$-PU matrix of size $2\times 4$ and sequence length $4$.
\end{example}

\subsection{\textbf{Extension of ZCZ Widths}}
In this subsection, we present a method to extend the ZCZ widths by using the seed ZPU matrices proposed in the previous subsection.

\begin{theorem}
 \label{Proposed:Construction:Theorem}
 Let $M$ and $K$ be two positive integers such that $K=MP$ for some positive integer $P$. Let $\bm{\mathcal{G}}(z)$ be a seed ZPU matrix of size $M\times K$, ZCZ width $Z$ and sequence length $L$ generated by (\ref{Seed:Z-PU:non:power:two}) and $\textbf{U}_M$ be any arbitrary $BH$ matrix of size $M\times M$. Then, an $M\times K$ matrix $\bm{\mathcal{G}}'(z)$ of polynomials with  unimodular coefficients is given by
\begin{equation}
\label{proposed:COnstruction:Larger:ZCZ}
\bm{\mathcal{G}}'(z)=\textbf{U}_M\cdot \textbf{D}_M(z)\cdot \bm{\mathcal{G}}(z^M),
\end{equation} where $\textbf{D}_M(z)=\text{diag}\Big(1,z^{-1}, \cdots, z^{-(M-1)}\Big)$. Then, the matrix  $\bm{\mathcal{G}}'(z)$ is an optimal unimodular $Z'$-PU matrix with size $M\times K$, sequence length $L'=ML$ and  ZCZ width $Z'=MZ$.
\end{theorem}
\begin{IEEEproof}
Let $\bm{\mathcal{G}}'(z)=\Big[g_{m\mu}'(z)\Big]_{M\times K}$, $\bm{\mathcal{G}}(z)=\left[g_{m\mu}(z)\right]_{M\times K}$ and $\textbf{U}_M=[u_{mn}]_{M\times M}$. Note that each sequence $g_{m\mu}(z)$ has length $L$. Since $\bm{\mathcal{G}}(z)$ is an $M\times K$ ZPU matrix with ZCZ width $Z$, we can write
\begin{align}
\label{ZPU:seed:Theorem}
S_{\textbf{g}_\mu,\textbf{g}_\nu}(z)={\color{blue}c\cdot \delta(\mu-\nu)}{\color{red}+\sum_{\tau=Z}^{L-1}S_{\textbf{g}_\mu,\textbf{g}_\nu}[\tau]\cdot z^{-\tau}},
\end{align} where $c$ is a positive constant describing the matrix energy. Also, we have $\textbf{u}^H_{n}\cdot \textbf{u}_{n'}=M\cdot \delta(n-n')$, where $\textbf{u}_{n}$ is the $n$-th column vector of $\textbf{U}_M$. Then, according to (\ref{proposed:COnstruction:Larger:ZCZ}), the $\mu$-th column of the matrix $\bm{\mathcal{G}}'(z)$ is given by
\begin{equation}
\textbf{g}'_\mu(z)=\Big[g'_{0\mu}(z),  g'_{1\mu}(z), \cdots,  g'_{(M-1)\mu}(z)\Big]^T,
\end{equation} where $0\leq \mu\leq K-1$ and the sequence $g_{m\mu}^{'}(z)$ of length $L'=ML$ can be written by
\begin{equation}
\label{entry_G:longer}
g_{m\mu}'(z)=\sum_{n=0}^{M-1}\sum_{l=0}^{L-1}u_{mn}\cdot g_{n\mu}[l]\cdot z^{-(n+Ml)},
\end{equation} where $m=0,1,\cdots, M-1$. Since $\textbf{U}_M$ has unimodular entries and $g_{n\mu}(z)$ is a unimodular sequence, the constructed sequence $g'_{m\mu}(z)$ is also a unimodular sequence of length $ML$.
Next, we calculate the sum of ACCFs between the $\mu$-th and $\nu$-th columns of the matrix $\bm{\mathcal{G}}'(z)$ as follows
\begin{align*}
&S_{\textbf{g}'_\mu,\textbf{g}'_\nu}(z)=\sum_{m=0}^{M-1}R_{\textbf{\textit{g}}'_{m\mu}, \textbf{\textit{g}}'_{m\nu}}(z) \nonumber \\
&=\sum_{m=0}^{M-1}g'_{m\mu}(z^{-1})\cdot g_{m\nu}^{'*}(z) \nonumber \\
&=\sum_{m=0}^{M-1}\sum_{n=0}^{M-1}\sum_{n'=0}^{M-1}\sum_{l=0}^{L-1}\sum_{l'=0}^{L-1}u_{mn}\cdot u_{mn'}^{*}\cdot g_{n\mu}[l]\cdot g_{n'\nu}^{*}[l']\cdot z^{-((n'-n)+M(l'-l))} \quad (\text{From (\ref{entry_G:longer})}) \nonumber \\
&=\sum_{n=0}^{M-1}\sum_{n'=0}^{M-1}\sum_{l=0}^{L-1}\sum_{l'=0}^{L-1}\textbf{u}^H_{n'}\cdot \textbf{u}_{n}\cdot g_{n\mu}[l]\cdot g_{n'\nu}^{*}[l']\cdot z^{-((n'-n)+M(l'-l))} \nonumber \\
&=\sum_{n=0}^{M-1}\sum_{n'=0}^{M-1}\sum_{l=0}^{L-1}\sum_{l'=0}^{L-1}M\cdot \delta(n-n')\cdot g_{n\mu}[l]\cdot g_{n'\nu}^{*}[l']\cdot z^{-((n'-n)+M(l'-l))} \quad (\because \textbf{U}^H_M\cdot \textbf{U}_M=M\textbf{I}_M) \nonumber \\
&=M\sum_{n=0}^{M-1}\sum_{l=0}^{L-1}\sum_{l'=0}^{L-1} g_{n\mu}[l]\cdot g_{n\nu}^{*}[l']\cdot z^{-M(l'-l)} \nonumber \\
\end{align*}
\begin{align}
&=M\sum_{n=0}^{M-1}\sum_{\tau=0}^{L-1}\sum_{l=0}^{L-\tau-1} g_{n\mu}[l]\cdot g_{n\nu}^{*}[l+\tau]\cdot z^{-M\tau}\quad (\text{Let $\tau=l'-l$}) \nonumber \\
&=M\sum_{n=0}^{M-1}\sum_{\tau=0}^{L-1}R_{\textbf{\textit{g}}_{n\mu},\textbf{\textit{g}}_{n\nu}}[\tau]\cdot z^{-M\tau} \nonumber \\
&=M\sum_{n=0}^{M-1}\left(\sum_{\tau=0}^{Z-1}R_{\textbf{\textit{g}}_{n\mu},\textbf{\textit{g}}_{n\nu}}[\tau]\cdot z^{-M\tau}{\color{red}+\sum_{\tau=Z}^{L-1}R_{\textbf{\textit{g}}_{n\mu},\textbf{\textit{g}}_{n\nu}}[\tau]\cdot z^{-M\tau}}\right) \nonumber \\
&={\color{blue}M\sum_{n=0}^{M-1}\sum_{\tau=0}^{Z-1}R_{\textbf{\textit{g}}_{n\mu},\textbf{\textit{g}}_{n\nu}}[\tau]\cdot z^{-M\tau}}{\color{red}+M\sum_{n=0}^{M-1}\sum_{\tau=Z}^{L-1}R_{\textbf{\textit{g}}_{n\mu},\textbf{\textit{g}}_{n\nu}}[\tau]\cdot z^{-M\tau}} \nonumber \\
&={\color{blue}M\sum_{\tau=0}^{Z-1}S_{\textbf{g}_\mu,\textbf{g}_\nu}[\tau]\cdot z^{-M\tau}}{\color{red}+M\sum_{\tau=Z}^{L-1}S_{\textbf{g}_\mu,\textbf{g}_\nu}[\tau]\cdot z^{-M\tau}} \nonumber \\
&={\color{blue}Mc\cdot \delta(\mu-\nu)}{\color{red}+M\sum_{\tau=Z}^{L-1}S_{\textbf{g}_\mu,\textbf{g}_\nu}[\tau]\cdot z^{-M\tau}}. \quad (\text{From (\ref{ZPU:seed:Theorem})})
\end{align} Thus, the matrix of ACCF sums is given by
\begin{align}
f_{Z'}\left(\widetilde{\bm{\mathcal{G}}'(z)}\cdot \bm{\mathcal{G}}'(z)\right)&=\Big[f_{Z'}\left(S_{\textbf{g}'_\mu,\textbf{g}'_\nu}(z)\right)\Big]_{K\times K} \nonumber \\
&= \Big[{\color{blue}Mc\cdot \delta(\mu-\nu)}\Big]_{K\times K} \nonumber \\
&=Mc\cdot \textbf{I}_K\ \text{for the ZCZ width $Z'=MZ$}.
\end{align} That is, $\bm{\mathcal{G}}'(z)$ is an $M\times K$ $Z'$-PU matrix with ZCZ width $Z'=MZ$ and sequence length $L'=ML$. Also, we have $K/M=\lfloor{L/Z}\rfloor=\lfloor{ML/MZ}\rfloor=\lfloor{L'/Z'}\rfloor$. So, $\bm{\mathcal{G}}'(z)$ is an optimal $Z'$-PU matrix. This completes the proof.
\end{IEEEproof}
\begin{remark}
We can also construct optimal ZPU matrix by using the output matrix as a seed matrix in \textit{Theorem \ref{Proposed:Construction:Theorem}}. The length $L'$ and ZCZ width  $Z'$ of the constructed sequences are increased proportionally by meeting the set size upper bound. We describe this by Table \ref{Table:Z-PU:parameter}.
\end{remark}
\begin{center}
\captionof{table}{Examples of Optimal ZPU Matrices}  \label{Table:Z-PU:parameter}
	%\resizebox{\linewidth}{!}{
		\renewcommand{\arraystretch}{1.2}
\begin{tabular}{|l|l|l|l|l|l|}
\hline
\multicolumn{1}{|c|}{$M$} & \multicolumn{1}{c|}{$K$} & \multicolumn{1}{c|}{$L$} & \multicolumn{1}{c|}{$Z$} & \multicolumn{1}{c|}{Phases} &\multicolumn{1}{c|}{$K/M=\lfloor{L/Z}\rfloor$} \\
\hline
\hline
\multicolumn{1}{|c|}{} & \multicolumn{1}{c|}{} & \multicolumn{1}{c|}{$4$} & \multicolumn{1}{c|}{$2$} &\multicolumn{1}{c|}{}& \multicolumn{1}{c|}{} \\
%\cline{3-4}
\multicolumn{1}{|c|}{$2$} & \multicolumn{1}{c|}{$4$} & \multicolumn{1}{c|}{$8$} & \multicolumn{1}{c|}{$4$} & \multicolumn{1}{c|}{$2$} & \multicolumn{1}{c|}{} \\
%\cline{3-4}
\multicolumn{1}{|c|}{} & \multicolumn{1}{c|}{} & \multicolumn{1}{c|}{$16$} & \multicolumn{1}{c|}{$8$} & \multicolumn{1}{c|}{}&\multicolumn{1}{c|}{} \\
%\cline{3-4}
\multicolumn{1}{|c|}{} & \multicolumn{1}{c|}{} & \multicolumn{1}{c|}{$32$} & \multicolumn{1}{c|}{$16$} & \multicolumn{1}{c|}{}& \multicolumn{1}{c|}{} \\
\cline{1-5}
\multicolumn{1}{|c|}{} & \multicolumn{1}{c|}{} & \multicolumn{1}{c|}{$6$} & \multicolumn{1}{c|}{$3$} &\multicolumn{1}{c|}{}& \multicolumn{1}{c|}{} \\
%\cline{3-4}
\multicolumn{1}{|c|}{$3$} & \multicolumn{1}{c|}{$6$} & \multicolumn{1}{c|}{$18$} & \multicolumn{1}{c|}{$9$} & \multicolumn{1}{c|}{$3$}& \multicolumn{1}{c|}{Optimal} \\
%\cline{3-4}
\multicolumn{1}{|c|}{} & \multicolumn{1}{c|}{} & \multicolumn{1}{c|}{$54$} & \multicolumn{1}{c|}{$27$} & \multicolumn{1}{c|}{}& \multicolumn{1}{c|}{} \\
%\cline{3-4}
\multicolumn{1}{|c|}{} & \multicolumn{1}{c|}{} & \multicolumn{1}{c|}{$162$} & \multicolumn{1}{c|}{$81$} & \multicolumn{1}{c|}{}&\multicolumn{1}{c|}{} \\
\cline{1-5}
\multicolumn{1}{|c|}{} & \multicolumn{1}{c|}{} & \multicolumn{1}{c|}{$32$} & \multicolumn{1}{c|}{$16$} & \multicolumn{1}{c|}{}& \multicolumn{1}{c|}{} \\
%\cline{3-4}
\multicolumn{1}{|c|}{$4$} & \multicolumn{1}{c|}{$8$} & \multicolumn{1}{c|}{$128$} & \multicolumn{1}{c|}{$64$} & \multicolumn{1}{c|}{$2$}& \multicolumn{1}{c|}{} \\
%\cline{3-4}
\multicolumn{1}{|c|}{} & \multicolumn{1}{c|}{} & \multicolumn{1}{c|}{$512$} & \multicolumn{1}{c|}{$256$} &\multicolumn{1}{c|}{}& \multicolumn{1}{c|}{} \\
%\cline{3-4}
\multicolumn{1}{|c|}{} & \multicolumn{1}{c|}{} & \multicolumn{1}{c|}{$2048$} & \multicolumn{1}{c|}{$1024$} & \multicolumn{1}{c|}{}&\multicolumn{1}{c|}{} \\
\hline
\end{tabular}
\end{center}
\begin{remark}
By applying (\ref{Seed:Z-PU:non:power:two}) and (\ref{proposed:COnstruction:Larger:ZCZ}), we can  construct optimal ZPU matrices with more flexible parameters using wide range of $BH$ matrices compared to the previous construction methods.
\end{remark}
We illustrate our proposed construction of optimal ZPU matrix with larger ZCZ width by the following example. We will give one example of a new $9$-PU matrix of size $3\times 6$ and sequence length $18$ with $3$-PSK constellation.
\begin{example}
Let $M=3$ and $K=6$ for $P=2$. Let $\bm{\mathcal{G}}(z)$ be a seed ZPU matrix with sequence length $L=6$ and ZCZ width $Z=3$ given in \textit{Example \ref{example:seed:Z-PU:Matrix:M3}} and $\textbf{U}_M=\textbf{F}_3$. Applying \textit{Theorem \ref{Proposed:Construction:Theorem}}, we can construct a $3\times 6$ $Z'$-PU matrix $\bm{\mathcal{G}}'(z)$ given by
\begin{align}
\bm{\mathcal{G}}'(z)&=\textbf{U}_M\cdot \textbf{D}_M(z)\cdot \bm{\mathcal{G}}(z^M)=\textbf{F}_3\cdot \textbf{D}_3(z)\cdot \bm{\mathcal{G}}(z^3),
\end{align} where $\textbf{D}_3(z)=\text{diag}\Big(1,z^{-1},z^{-2}\Big)$. Since the matrix $\bm{\mathcal{G}}(z)$ has sequences with $3$-PSK constellation, the matrix $\bm{\mathcal{G}}'(z)$ consists of sequences with $3$-PSK constellation. The ZCZ width is $Z'=MZ=3\cdot 3=9$ and sequence length is $L'=ML=3\cdot 6=18$.
   The matrix of ACCF sums is given by
\begin{align}
\widetilde{\bm{\mathcal{G}}'(z)}\cdot\bm{\mathcal{G}}'(z)&=\Big[S_{\textbf{g}'_\mu,\textbf{g}'_\nu}(z)
\Big]_{6\times 6}=\left[{\color{blue}54\cdot \delta(\mu-\nu)}{\color{red}+\sum_{9\leq |\tau|< 18}S_{\textbf{g}'_\mu,\textbf{g}'_\nu}[\tau]\cdot z^{-\tau}}\right]_{6\times 6}.
\end{align} Therefore,  we have $f_{Z'}\left(\widetilde{\bm{\mathcal{G}}'(z)}\cdot\bm{\mathcal{G}}'(z)\right)=54\cdot \textbf{I}_6$ for the ZCZ width $Z'=9$. Also, we have $K/M=P=\lfloor{L'/Z'}\rfloor$ and hence $\bm{\mathcal{G}}'(z)$ is an optimal unimodular $9$-PU matrix of size $3\times 6$ and sequence length $18$ with $3$-PSK constellation. We have written out this $9$-PU matrix by Table \ref{Table:Z-PU:M3:K6:Z9:L18} in which only exponents of $\omega=e^{-2\pi\sqrt{-1}/3}$ are given.
\end{example}

According to \textit{Theorem \ref{Proposed:Construction:Theorem}}, we have the following corollary for binary sequences.
\begin{corollary}
\label{corollary:theorem3}
Let $\textbf{H}_{2^{m+k}}$ and $\textbf{H}_{2^{m}}$ be two binary Hadamard matrices of size $2^{m+k}\times 2^{m+k}$ and $2^{m}\times 2^{m}$, respectively. Let $\bm{\mathcal{G}}(z)$ be an optimal binary $2^m$-PU matrix of size $2^{m}\times 2^{m+k}$ and sequence length $2^{m+k}$ constructed from \textit{Corollary \ref{cor:theorem2}}. Then, the matrix
$\bm{\mathcal{G}}'(z)$ given by (\ref{proposed:COnstruction:Larger:ZCZ}) is an optimal binary $2^{2m}$-PU matrix of size $2^{m}\times 2^{m+k}$ and sequence length $2^{2m+k}$. That is, $\bm{\mathcal{G}}'(z)$ represents an optimal binary $(2^{m+k},2^{2m})$-ZCCS$^{2^{2m+k}}_{2^{m}}$.
\end{corollary}
The next example demonstrates the above corollary.
\begin{example}
Let $M=2$ and $K=4$ for $P=2$. Let $\bm{\mathcal{G}}(z)$ be a binary seed ZPU matrix with sequence length $L=4$ and ZCZ width $Z=2$ given in \textit{Example \ref{example:seed:Z-PU:Matrix}} and $\textbf{H}_2$ be a binary Hadamard matrix of size $2\times 2$. Applying \textit{Theorem \ref{Proposed:Construction:Theorem}}, we can construct a binary $Z'$-PU matrix $\bm{\mathcal{G}}'(z)$ of size $2\times 4$ given by
\begin{align}
\bm{\mathcal{G}}'(z)&=\textbf{H}_M\cdot \textbf{D}_M(z)\cdot \bm{\mathcal{G}}(z^M)=\textbf{H}_2\cdot \textbf{D}_2(z)\cdot \bm{\mathcal{G}}(z^2) \nonumber \\
&\equiv \begin{bmatrix}
+++-+++-&++-+++-+&+++----+&++-+--+- \\
+-+++-++&+---+---&+-++-+--&+----+++
\end{bmatrix},
\end{align} where $\textbf{D}_2(z)=\text{diag}(1,z^{-1})$. The ZCZ width is $Z'=MZ=2\cdot 2=4$ and sequence length is $L'=ML=2\cdot 4=8$.  The matrix of ACCF sums is given by
\begin{align}
\widetilde{\bm{\mathcal{G}}'(z)}\cdot\bm{\mathcal{G}}'(z)&=\Big[S_{\textbf{g}'_\mu,\textbf{g}'_\nu}(z)
\Big]_{4\times 4}=\left[{\color{blue}16\cdot \delta(\mu-\nu)}{\color{red}+\sum_{4\leq |\tau|<8}S_{\textbf{g}'_\mu,\textbf{g}'_\nu}[\tau]\cdot z^{-\tau}}\right]_{4\times 4}.
\end{align} Thus, $f_{Z'}\left(\widetilde{\bm{\mathcal{G}}'(z)}\cdot\bm{\mathcal{G}}'(z)\right)=16\cdot \textbf{I}_4$ for the ZCZ width $Z'=4$. Also, we have $K/M=2=\lfloor{L'/Z'}\rfloor$ and hence $\bm{\mathcal{G}}'(z)$ is an optimal binary $4$-PU matrix of size $2\times 4$ and sequence length $8$.
\end{example}
 \begin{center}
\captionof{table}{A New $3\times 6$ Optimal $9$-PU Matrix with Sequence Length $18$}  \label{Table:Z-PU:M3:K6:Z9:L18}
	%\resizebox{\linewidth}{!}{
		\renewcommand{\arraystretch}{1.05}
\begin{tabular}{|l|l||l|l||l|l|}
\hline
 & $0     0     0     0     1     2     0     2     1     0     0     0     0     1     2     0     2     1$ &  & $0     0     0     0     1     2     1     0     2     1     1     1     2     0     1     2     1     0$ &  & $0     0     0     1     2     0     0     2     1     2     2     2     2     0     1     1     0     2$ \\
$\textbf{x}_0$ & $0     1     2     0     2     1     0     0     0     0     1     2     0     2     1     0     0     0$ & $\textbf{x}_1$ & $0     1     2     0     2     1     1     1     1     1     2     0     2     1     0     2     2     2$ & $\textbf{x}_2$ & $0     1     2     1     0     2     0     0     0     2     0     1     2     1     0     1     1     1$ \\
 & $0     2     1     0     0     0     0     1     2     0     2     1     0     0     0     0     1     2$ &  & $0     2     1     0     0     0     1     2     0     1     0     2     2     2     2     2     0     1$ &  & $ 0     2     1     1     1     1     0     1     2     2     1     0     2     2     2     1     2     0$ \\
\hline
\hline
 & $0     0     0     1     2     0     2     1     0     0     0     0     1     2     0     2     1     0
$ &  & $0     0     0     2     0     1     2     1     0     1     1     1     0     1     2     1     0     2$ &  & $ 0     0     0     2     0     1     1     0     2     2     2     2     1     2     0     0     2     1$ \\
$\textbf{x}_3$ & $0     1     2     1     0     2     2     2     2     0     1     2     1     0     2     2     2     2$ & $\textbf{x}_4$ & $ 0     1     2     2     1     0     2     2     2     1     2     0     0     2     1     1     1     1$ & $\textbf{x}_5$ & $0     1     2     2     1     0     1     1     1     2     0     1     1     0     2     0     0     0$ \\
 & $0     2     1     1     1     1     2     0     1     0     2     1     1     1     1     2     0     1$ &  & $0     2     1     2     2     2     2     0     1     1     0     2     0     0     0     1     2     0$ &  & $0     2     1     2     2     2     1     2     0     2     1     0     1     1     1     0     1     2$ \\
\hline
\end{tabular}
\end{center}
\section{\textbf{Conclusion}}
 \label{Z-PU:sec:conclusion}
In this paper, we have introduced a new concept, called ZPU matrix, whose corresponding ZCCS has zero (nontrivial) aperiodic correlation sums at all time-shifts ranging from $-(Z-1)$ to $(Z+1)$, where $Z$ denotes the ZCZ width which may be less than the sequence length $L$.  We have shown that an $M\times K$ ZPU matrix exists when $K\leq M \lfloor{L/Z}\rfloor$, in contrast to an existing PU matrix satisfying $K\leq M$. We have developed a unifying construction framework for optimal ZPU matrices, which includes existing PU matrices as a special case. Our proposed construction framework offers flexible ZCCS parameters compared to the previous construction methods.

\section*{\textbf{Acknowledgment}}
Shibsankar Das and Zilong Liu are deeply indebted to Dr. Srdjan Budi\v{s}in at the RT-RK, Novi Sad, Serbia, whose insightful thoughts inspired this work.

\bibliographystyle{IEEEtran}
\bibliography{refs}

\end{document}